\documentclass[aoas]{imsart}

\RequirePackage{amsthm,amsmath,amsfonts,amssymb}
\RequirePackage[authoryear]{natbib}
\RequirePackage[colorlinks,citecolor=blue,urlcolor=blue]{hyperref}
\RequirePackage{graphicx}
\usepackage{booktabs}
\usepackage{float}
\usepackage{algorithm2e}
\DeclareMathOperator*{\argmin}{arg\,min}
\startlocaldefs
\theoremstyle{plain}

\newtheorem{theorem}{Theorem}[section]

\theoremstyle{remark}
\newtheorem{rem}[theorem]{Remark}

\newcommand{\beginsupplement}{%
        \setcounter{table}{0}
        \renewcommand{\thetable}{S\arabic{table}}%
        \setcounter{figure}{0}
        \renewcommand{\thefigure}{S\arabic{figure}}%
     }

\endlocaldefs

\begin{document}

\begin{frontmatter}
\title{An Effective Method for Identifying\\ Clusters of Robot Strengths}
\runtitle{Identifying Clusters of Robot Strengths}

\begin{aug}
\author[A]{\fnms{Jen-Chieh}~\snm{Teng}\ead[label=e1]{D09948011@ntu.edu.tw}},
\author[B]{\fnms{Chin-Tsang}~\snm{Chiang}\ead[label=e2]{chiangct@ntu.edu.tw}\orcid{0000-0001-7957-7061}}
\and
\author[C,D]{\fnms{Alvin}~\snm{Lim}\ead[label=e3]{alvin.lim@emory.edu}\orcid{0000-0001-7886-5643}}
\address[A]{Data Science Degree Program, National Taiwan University \printead[presep={,\ }]{e1}}
\address[B]{Institute of Applied Mathematical Sciences, National Taiwan University\printead[presep={,\ }]{e2}}
\address[C]{Data Science Department, Measured, Inc.\printead[presep={\ }]{}}
\address[D]{Goizueta Business School, Emory University\printead[presep={,\ }]{e3}}
\end{aug}

\begin{abstract}
In the analysis of qualification stage data from FIRST Robotics Competition (FRC) championships, the ratio (1.67 -- 1.68) of the number of observations (110 -- 114 matches) to the number of parameters (66 -- 68 robots) in each division has been found to be quite small for the most commonly used winning margin power rating (WMPR) model.
This usually leads to imprecise estimates and inaccurate predictions in such three-on-three matches that FRC tournaments are composed of.
With the recognition of a clustering feature in estimated robot strengths, a more flexible model with latent clusters of robots was proposed to alleviate overparameterization of the WMPR model.
Since its structure can be regarded as a dimension reduction of the parameter space in the WMPR model, the identification of clusters of robot strengths is naturally transformed into a model selection problem.
Instead of comparing a huge number of competing models $(7.76\times 10^{67}$ to $3.66\times 10^{70})$, we develop an effective method to estimate the number of clusters, clusters of robots and robot strengths in the format of qualification stage data from the FRC championships.
The new method consists of two parts: (i) a combination of hierarchical and non-hierarchical classifications to determine candidate models; and (ii) variant goodness-of-fit criteria to select optimal models.
In contrast to existing hierarchical classification, each step of our proposed non-hierarchical classification is based on estimated robot strengths from a candidate model in the preceding non-hierarchical classification step.
A great advantage of the proposed methodology is its ability to consider the possibility of reassigning robots to other clusters.
To reduce overestimation of the number of clusters by the mean squared prediction error criteria, corresponding Bayesian information criteria are further established as alternatives for model selection.
With a coherent assembly of these essential elements, a systematic procedure is presented to perform the estimation of parameters. 
In addition, we propose two indices to measure the nested relation between clusters from any two models and monotonic association between robot strengths from any two models.
Data from the 2018 and 2019 FRC championships and a simulation study are also used to illustrate the applicability and superiority of our proposed methodology.
\end{abstract}


\end{frontmatter}

\begin{section}{Introduction}

The FIRST Robotics Competition (FRC) is a sports-like robotic competition sponsored by the international youth organization For Inspiration and Recognition of Science and Technology (FIRST). According to the 2022 FRC Game Manual \citep{manual2022:first}, ``FRC combines the excitement of sport with the rigors of science and technology. Teams of students are challenged to design, build and program industrial-size robots and compete for awards, while they also create a team identity, raise funds, hone teamwork skills and advance respect and appreciation for STEM within the local community.'' \cite{wiki:first} provides a more detailed description of the FIRST organization and its history.

Each year's FRC season begins with a kickoff event in January when a new and challenging game is introduced with specific rules and mechanics. Teams of high school students around the world then design and build a robot to participate in local, regional and international tournaments. While a new game is introduced each year, the format of a match remains constant over the years. A match involves an alliance of three robots from three teams going against another alliance of three other robots from three other teams in a game defined by the rules and mechanics of the year. Winning a match requires an alliance to score more points over the opposing alliance according to game rules and mechanics. The opposing alliances in a match are randomly designated as a ``blue alliance'' and a ``red alliance'' and robots representing each alliance will have bumpers of their alliance's color
(see Figure \ref{plot:2019field} for visual representation of the game field).

\begin{figure}[htbp]
\centering
 \includegraphics[scale=0.275]{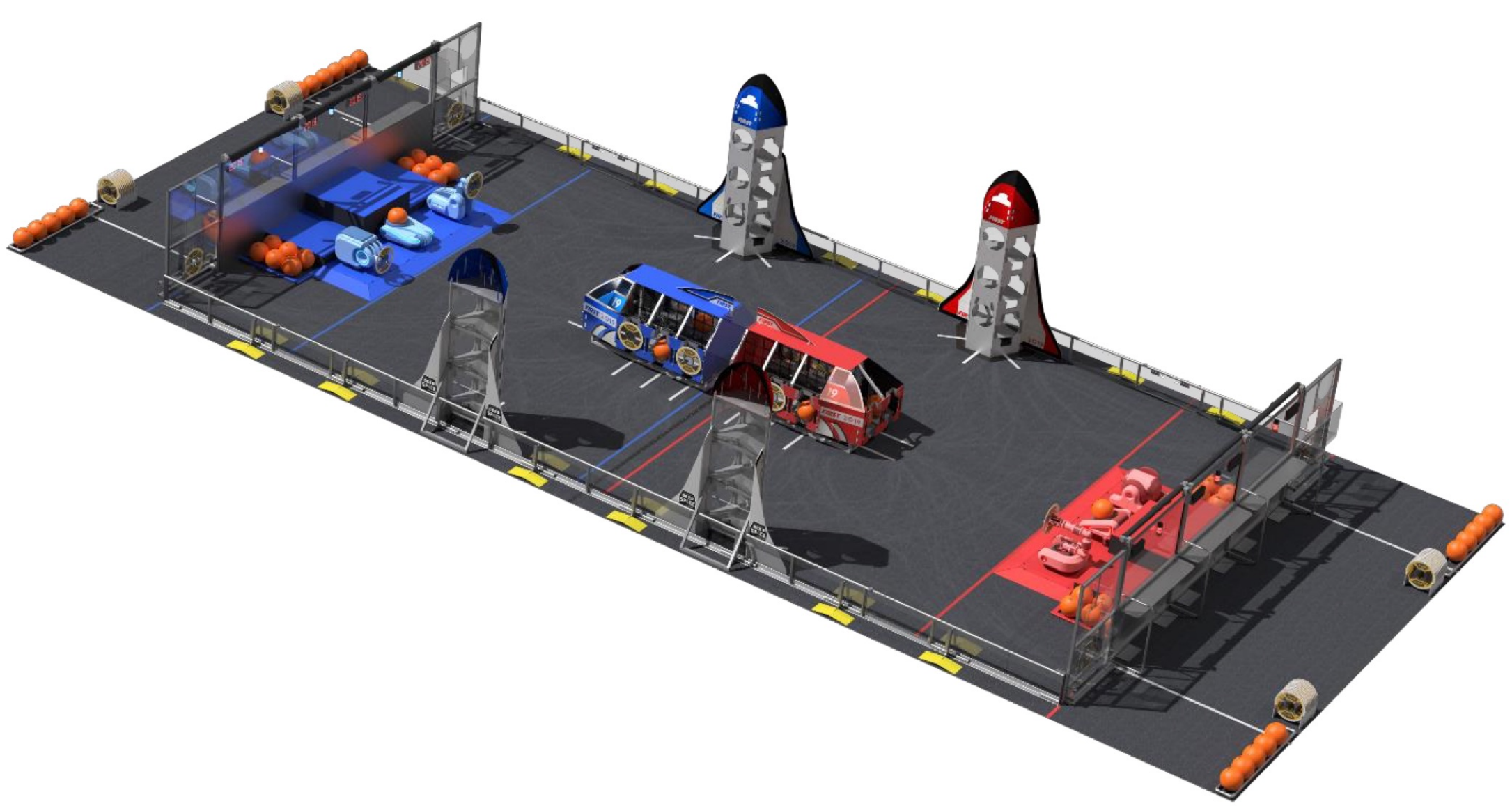}
 \caption{{\footnotesize The 2019 FRC game field. Source: \cite{manual:first}.}} \label{plot:2019field}
\end{figure}

All local, regional and world championship tournaments are divided into a qualification stage and a play-off stage. In the qualification stage, the participating teams’ robots are assigned by an algorithm into a predetermined number of matches and in each match, a robot is randomly assigned to one of two opposing alliances. After each match, the three robots in an alliance are assigned the same number of ranking points (different from scores earned in the match) according to game rules and the outcome of the match. At the end of the qualification stage, a predetermined number of robots with the most ranking points then qualify for the playoff stage. The robots' average scores across qualification matches are used as an auxiliary measure to break ties when necessary.     

In contrast to the qualification stage, where robots are randomly assigned alliances for each match, a predetermined number of top teams whose robots qualified for the playoff stage are asked to form their own alliances of four robots each. These alliances remain throughout matches in the playoff stage, where alliances attempt to eliminate each other across matches. As in the qualification stage, each match in the playoff stage will involve three robots each from opposing alliances with one robot each in reserve. Since each robot is identified with the team that built and owns the robot, we will be using the terms robot and team (or robotics team) interchangeably henceforth. More detailed descriptions of FRC games, matches and tournaments can be found in \cite{LCT2021}.

On the basis of very limited official statistics (regular scores, bonuses and penalties), one of the most challenging tasks for the teams when forming alliances for the playoff stage is to accurately assess the true ``strengths'' of robots based on their contributions in winning or losing matches. There is still no reliable robot rating/ranking system, albeit the FRC and robotics teams might have their own measures or models to quantify the strengths of competing robots. For example, as mentioned earlier, official rankings of robotics teams are determined by a system of ranking points with the average scores as an auxiliary measure to break ties when necessary. However, a strong (weak) robot may be underrated (overrated) by both the measures since the same number of ranking points and the same score is assigned to all robots in an alliance after a qualification match regardless of each individual robot's actual contribution to the performance of the alliance. 

Due to the importance of objectively assessing a robot's potential contribution to an alliance in the playoff stage of a tournament and the absence of such accurate and reliable assessment measures beyond those proposed in \cite{LCT2021}, we devote this study to building a more accurate model of robot strengths and providing a more effective estimation of model parameters. For the remainder of this section, we review existing rating models that are currently in use by robotics teams and traditional sports teams.

In robotics forums, the calculated contribution to winning margin (CCWM), offensive power rating (OPR) and winning margin power rating (WMPR) models have been popularly used to assess robot strengths.
As shown by \cite{law:2008}, the CCWM model is a special case of the WMPR model and is meaningless in application.
Based on the calculated contribution measure of Karthik Kanagasabapthy in 2004, \cite{tba:opr} modeled the average alliance score as the sum of robot strengths and proposed a least squares estimation procedure.
Before this work, \cite{weingart:2006} termed such a measure as the OPR and explained its detailed computation.
Due to ignoring actions of robots (as quantified by final match scores) in the opposing alliance, the OPR model tends to have biased estimates.
By considering an interaction between alliances and latent factors in each match, \cite{gardner:2015} proposed the WMPR model to characterize the effects of robot strengths on the difference in final match scores.
Similar to this model, the adjusted plus/minus rating model and its generalizations have been found to be useful for paired comparison data in traditional sports and esports \citep[cf.][among others]{Stefani1977, Stefani1980, CN1995, rosenbaum:2004, macdonald:2011, schuckers:2011, saebo:2015, hass:2018, hvattum:2019, clark:2020}.
In practice, there is no need to introduce a home-court advantage and clutch time/garbage time play in the WMPR model for the FRC.

In an application to qualification stage data from the 2018 FRC Houston and Detroit championships, \cite{LCT2021} demonstrated the poor performance of the WMPR model in estimating robot strengths and predicting match scores (or outcomes).
This can be explained by a relatively large number of parameters (number of robots) as compared to the number of observations (number of matches) in each division.
With the recognition of a clustering feature in estimated robot strengths, the WMPRC model, which is an extension of the WMPR model with the consideration of latent clusters of robots, was proposed by \cite{LCT2021} to alleviate the overparameterization problem.
The number of candidate cluster regression functions is huge for each given number of clusters, albeit the WMPRC model has the form of clusterwise linear regression of \cite{S1979}. Therefore, the high computational complexity of the conditional mixture method of \cite{DC1998} makes the estimation of robot strengths challenging. In the format of qualification stage data, there is also no advantage in introducing a pairwise fusion penalty method in the spirit of \cite{bondell2008simultaneous}. By applying the central linkage clustering method of \cite{SM1958} on estimated robot strengths from the WMPR model, \cite{LCT2021} developed a simple method to determine WMPRC candidate models for different numbers of clusters.
However, the performance of these competing models is highly related to imprecise initial estimates.
To reduce this influence, the authors improved their first method by sequentially merging two clusters based on estimated robot strengths from a WMPRC candidate model determined in the preceding hierarchical classification step.
Two types of mean squared prediction error (MSPE) criteria were further established to select optimal models from these WMPRC candidate models.
As shown in the simulation study, the MSPE measures are underestimated by their cross-validation estimates.
Even though estimated MSPE curves exhibit the U-shaped patterns, the number of clusters is generally overestimated by their minimizers.

For the hierarchical classification in the second method of \cite{LCT2021}, robots may be classified into incorrect clusters.
To tackle this problem, a non-hierarchical classification is proposed to examine the possibility of reassigning robots to other clusters. 
With the determined clusters of robots in each hierarchical classification step, we fit all possible WMPRC models with a robot in a new cluster and the rest robots in the original clusters.
The clusters of robots are further updated by applying the central linkage clustering method on estimated robot strengths corresponding to new clusters in these fitting WMPRC models.
This process is repeated until updated clusters of robots are the same as those in the preceding iterations.
Different from existing hierarchical classifications, some steps of our hierarchical classification are based on estimated robot strengths from WMPRC candidate models in their preceding non-hierarchical classification steps.
To reduce overestimation of the number of clusters by the MSPE criteria, the corresponding MSPE-based Bayesian information (MSPEB) criteria are established as alternatives to select optimal models from WMPRC candidate models.
The conducted simulation shows that the performance of the MSPE criteria can be enhanced by their MSPEB criteria when the proportion of sufficiently separated robot strengths, which are assessed by the ratios of pairwise differences between different robot strengths to the standard deviation of the error, is higher or lower than some threshold value.
Briefly speaking, the developed method is formed by two parts: one is a combination of hierarchical and non-hierarchical classifications and the other is any one of the MSPE and MSPEB criteria.
This new method generally outperforms its competitors in the format of qualification stage data from the FRC championships.
In this study, we also propose an index to measure the nested relation between clusters from any two models and modify the rank correlation of \cite{H1987} to measure the monotonic association between robot strengths from any two models.

The rest of this article is organized as follows. 
Section \ref{sec:WMPRC} presents the WMPRC model and reviews existing estimation methods.
An effective method is developed in Section \ref{sec:new} to estimate the model parameters.
In Section \ref{sec:analysis}, the methodology is further applied to qualification stage and playoff stage data from the 2018 and 2019 FRC Houston and Detroit championships.
Moreover, a simulation is conducted to assess and compare the performance of our method and its competitors.
Section \ref{sec:concluding} summarizes the achievements and findings and discusses future research directions.
\end{section}

\begin{section}{WMPRC Model and Existing Estimation Methods}  \label{sec:WMPRC}
We define the following notations. Let $Y_s$ denote the difference in scores between the red and blue alliances in match $s$; $D_s$ a binary outcome (1=win, 0=loss) in match $s$; $O_i$ robot $i$; and
$x_{si}$ the designated covariate of $O_i$ in match $s$ taking a value of $1$ if $O_i$ is assigned to the red alliance, $-1$ if $O_i$ is assigned to the blue alliance and $0$ otherwise, $s = 1, \dots, M, i = 1, \dots, K$.

\begin{subsection}{Model Structure} \label{subsec:model}


Let $c_o$ and $g^o = \big(g_{1}^o, \dots, g_{K}^o\big)^{\top}$, respectively, represent the corresponding number of clusters and clusters of robots in the WMPRC model with $g_{1}^o, \dots, g_{K}^o \in \{ 1, \dots, c_o \}$.
The relation between the difference in scores and robot strengths in this model is specified as follows:
\begin{align}
Y = X \beta^o + \varepsilon, \label{eq:lnmodel}
\end{align}
where $Y = (Y_1, \dots, Y_M)^{\top}$ is an $M \times 1$ response vector, $X = (x_{si})$ is an $M \times K$ covariate matrix, $\beta^o = \big(\beta^o_{g_{1}^o}, \dots, \beta^o_{g_{K}^o}\big)^{\top}$ is a $K \times 1$ parameter vector of robot strengths and $\varepsilon = (\varepsilon_1, \dots, \varepsilon_M)^{\top}$ is an $M \times 1$ vector of independent and identically distributed random errors from an unknown distribution function $F^o(\cdot)$ with mean zero and variance $\sigma^2$.
It is noted that the WMPR model is in the form of model (\ref{eq:lnmodel}) with
$c_o = K$.
Since $X$ is not a full-rank matrix, the constraint $\sum_{i=1}^K \beta^o_{g_{i}^o} = 0$ is imposed as most existing paired comparison models \citep[see][]{BT1952, B1953, CVF2013}.
The regression coefficients $\beta^o_{g_{1}^o}, \dots, \beta^o_{g_{K}^o}$ are, thus, interpreted as the relative robot strengths.
Just like in application to traditional sports, the primary research interest in the rating of robots in the FRC lies in predicting a match outcome.
It follows from model (\ref{eq:lnmodel}) that the probability of $\{ D_s = 1\}$, given $x_s = (x_{s1}, \dots, x_{sK})^{\top}$, can be expressed as
\begin{align}
P(D_s = 1|x_s) = 1 - F^o \left(-x_s^{\top}\beta^o \right), s = 1, \dots, M. \label{eq:condprob}
\end{align}
Since no assumption is made on the functional form of $F^o$, the Bradley-Terry model \citep{BT1952, B1953} and the Thurstone-Mosteller model \citep{T1927, M1951} are special cases of the above binary response model.

In fact, the perspective of the WMPRC model can be adopted to classify players into same skill levels in the United States Chess Federation (USCF) and the National Wheelchair Basketball Association (NWBA).
In contrast with the rating systems in these applications, the number of clusters and clusters of robots in the WMPRC model are unknown in advance of data analysis.
Furthermore, it may be inappropriate to determine the clusters of robots with some degrees of subjectivity.
As shown by \cite{LCT2021}, the estimation of $\beta^{o}$ can be transformed into model selection from WMPRC candidate models of the form:
\begin{align}
Y = X \beta^c + \varepsilon \text{ with } g^c, \label{eq:lnmodel_c}
\end{align}
where $\beta^c = \big(\beta^c_{g_{1}^c}, \dots, \beta^c_{g_{K}^c}\big)^{\top}$ is a $K \times 1$ parameter vector of robot strengths, $g^c = \big(g_{1}^c, \dots$, $g_{K}^c \big)^{\top}$ is a $K \times 1$ vector of clusters of robots with $g_{1}^c, \dots, g_{K}^c$ $\in \{ 1, \dots, c \}$ and $\varepsilon_1, \dots, \varepsilon_M$ in $\varepsilon$ are from an unknown distribution $F^c(\cdot)$, $c = 2, \dots, K$.
The probability of $\{ D_s = 1 \}$, given $x_s$, is directly derived as
\begin{align}
P(D_s = 1|x_s) = 1 - F^c \left(-x_s^{\top}\beta^c \right), s = 1, \dots, M. \label{eq:condprob_c}
\end{align}
\end{subsection}

\begin{subsection}{Classification and Model Selection} \label{section:em}
Although the estimation of parameters in (\ref{eq:lnmodel}) and (\ref{eq:condprob}) can be transformed into a model selection problem, it is impractical to fit a huge number of models $\big( \sum_{c=1}^K  \sum_{j=0}^{c-1} (-1)^j \binom{c}{j}(c-j)^K/c!\big)$ in the forms of (\ref{eq:lnmodel_c}) and (\ref{eq:condprob_c}).
Especially, this requires much higher computational cost and longer computational time.
To overcome such a limitation, \cite{LCT2021} proposed two simple methods to estimate the parameters $c_{o}$, $g^{o}$ and $\beta^{o}$.
For a WMPRC candidate model $Y=X\beta^{c}+\varepsilon$ with $g^c$, the authors 
estimated $\beta^c$ with the LSE $\widehat{\beta}^c$ and $F^c(\cdot)$ with the empirical distribution function $\widehat{F}^c(\cdot)$ of the residuals $e^c_{s} = Y_s - x_s^{\top} \widehat{\beta}^c, s = 1, \dots, M$.

Let $Y_0$ be a future observation of the difference in scores on a given covariate vector $x_0$, which is defined as $x_1, \dots, x_M$, with the corresponding $D_0$ for a match outcome.
For the predictors $\widehat{Y}_0 = x_0^{\top} \widehat{\beta}^c$, $\widehat{p}_0 = 1 - \widehat{F}^c \big(- \widehat{Y}_0\big)$ and $\widehat{D}_0 = I\big(\widehat{p}_0 
> 0.5 \big) + 0.5 I\big(\widehat{p}_0 =0.5 \big)$, their predictive capacities can be assessed by
\begin{align}
\textsc{mspe}_{\textsc{y}}\big(c,g^c\big)  &= E \Big[\left(Y_0 -  \widehat{Y}_0 \right)^2 \Big], \textsc{mspe}_{\textsc{p}}\big(c,g^c\big)  = E \Big[\left(D_0 - \widehat{p}_0 \right)^2 \Big], \text{ and} \label{eq:mspey} \\
\textsc{mspe}_{\textsc{d}}\big(c,g^c\big)  &= E \Big[\left(D_0 -  \widehat{D}_0 \right)^2 \Big],\text{ respectively}.\nonumber
\end{align}
It is noted that the measure $(1 - \textsc{mspe}_{\textsc{d}}(c,g^c))$ is an alternative expression for 
the probability of correct prediction (PCP), i.e.
\begin{align}
\textsc{pcp}\big(c,g^c\big) = & P\big(  \big(\widehat{p}_0  - 0.5\big) Y_0 > 0\big) + 0.5 P\big( \big(\widehat{p}_0   - 0.5\big) Y_0 = 0 \big).  \label{eq:pcp}
\end{align}
Based on qualification stage data $\{(Y_s, x_s)\}_{s=1}^M$, the following cross-validation estimators are naturally proposed for the corresponding MSPE measures in (\ref{eq:mspey}):
\begin{align}
&\widehat{\textsc{mspe}}_{\textsc{y}}\big(c,g^c\big)  =  \frac{1}{M}\sum_{s=1}^M 
\left(Y_s - \widehat{Y}_{\text{-}s} \right)^2,\label{eq:mspey_hat}
\widehat{\textsc{mspe}}_{\textsc{p}}\big(c,g^c\big)  =  \frac{1}{M}\sum_{s=1}^M 
\left(D_s - \widehat{p}_{\text{-}s} \right)^2,\text{ and}  \\ 
&\widehat{\textsc{mspe}}_{\textsc{d}}\big(c,g^c\big)  =  \frac{1}{M}\sum_{s=1}^M 
\left(D_s - \widehat{D}_{\text{-}s} \right)^2, c = 2, \dots, K, \nonumber
\end{align}
where $\widehat{Y}_{\text{-}s} = x_s^{\top} \widehat{\beta}_{\text{-}s}^c$, $\widehat{p}_{\text{-}s} = 1 - \widehat{F}_{\text{-}s}^c \big(- \widehat{Y}_{\text{-}s}\big)$, $\widehat{D}_{\text{-}s} = I\big(\widehat{p}_{\text{-}s} - 0.5 > 0\big) + 0.5 I\big(\widehat{p}_{\text{-}s} - 0.5 = 0\big)$ and $\big(\widehat{\beta}_{\text{-}s}^c, \widehat{F}_{\text{-}s}^c \big)$ is computed as $\big(\widehat{\beta}^c, \widehat{F}^c \big)$ with $(Y_s, x_s)$ excluded, $s = 1, \dots, M$.
From the context of regression analysis, these counterparts can be directly obtained by using the algebraic relation between $\widehat{\beta}_{\text{-}s}^c$ and $\big\{ \widehat{\beta}^c, X, e^c_{s} \big\}$ for each $s$.

For the estimation of the WMPRC model, \cite{LCT2021} applied the centroid linkage clustering method of \cite{SM1958} on the LSE $\widehat{\beta}^K$ of $\beta^K$ in the WMPR model $Y=X\beta^K + \varepsilon$ to determine the clusters of robots ${g}^c$ in $Y=X\beta^c + \varepsilon$ for each $c$.
An optimal model is further selected from these WMPRC candidate models with $ \max_{c} \widehat{\textsc{pcp}}(c,{g}^c)~( = 1 -  \min_{c} \widehat{\textsc{mspe}}_{\textsc{d}}(c,{g}^c))$ or $\min_{c} \widehat{\textsc{mspe}}_{\textsc{y}}(c,{g}^c)$. 
In practice, the estimated function $\widehat{\textsc{mspe}}_{\textsc{p}}(c,{g}^c)$ can be used as another criterion for model selection.
To avoid over-relying on an imprecise estimator $\widehat{\beta}^K$, the first method was improved by sequentially determining the clusters of robots ${g}^c$ in $Y = X\beta^{c} + \varepsilon$ based on the LSE $\widehat{\beta}^{c+1}$ of $\beta^{c+1}$ in $Y = X\beta^{c+1} + \varepsilon$ with ${g}^{c+1}$ for $c$ from $K-1$ to 2.
In an application to playoff data from the 2018 FRC Houston and Detroit championships, the second method was found to have better performance in the prediction of match scores and outcomes.
The data and R codes for both the methods are also available at \url{http://www.runmycode.org/companion/view/4012}.
\end{subsection}

\begin{rem}
Given the number of clusters $c_o$, the WMPRC model has the form of clusterwise linear regression of \cite{S1979}.
With a specified probability density function $f^o(\cdot)$ of the error $\varepsilon$, the unconditional mixture likelihood function of qualification stage data in each division of the FRC championships can be derived as follows:
\begin{align}
    &L\big( \beta^o_{1}, \dots, \beta^o_{c_o}, f^o, p_1, \dots, p_{c_o}\big) \label{eq:uncond} \\
    = & \sum_{g_1 = 1}^{c_o} \dots  \sum_{g_K = 1}^{c_o} \left( \prod_{s=1}^M f^o\left(Y_s - \beta^o_{g_1}x_{s1} - \dots - \beta^o_{g_K} x_{sK}\right) \prod_{k=1}^{c_o} p_k^{\sum_{i=1}^K I(g_i = k)} \right),\nonumber
\end{align}
where $p_k \in (0,1)$ is the probability of $\{g_i^o =k \}$ for each $i = 1, \dots, M$, $k = 1, \dots, c_o$, with the constraint $\sum_{k=1}^{c_o} p_k = 1$.
However, the maximum likelihood method is unsuitable for estimating robot strengths due to the unknown $c_o$ and $f^o$ in the proposed model and
the computational complexity in the maximization of (\ref{eq:uncond}). \qed
\end{rem}
\end{section}

\begin{section}{New Estimation Method} \label{sec:new}
We first introduce the essential elements of the developed method for identifying clusters of robot strengths.
A systematic estimation procedure and indices of nested relation and monotonic association are further proposed in the succeeding subsections.
Given clusters of robots $g^c$, the $k$th cluster $\{ O_i: g^c_i = k, i = 1, \dots, K \}$ is hereinafter denoted by $G_{k}^c$, $k = 1, \dots, c$.

\begin{subsection}{Background} \label{section:back}
In the second method of \cite{LCT2021}, which is referred to as the LCT method, the clusters of robots are sequentially updated based on estimated robot strengths from a WMPRC candidate model determined in the preceding hierarchical classification step.
For each WMPRC model determined by our hierarchical classification, which is different from that in the LCT method, a non-hierarchical classification is further proposed to examine the possibility of reassigning robots to other clusters.
As alternatives to the MSPE criteria, the MSPEB criteria are also established to select optimal models from WMPRC candidate models determined by the proposed classification system, a combination of hierarchical and non-hierarchical classifications.

In contrast with the hierarchical classification in the LCT method, our hierarchical classification is performed on the WMPR model $Y = X \beta^K + \varepsilon$ and WMPRC models $Y = X \beta^{c+1} + \varepsilon$ with $g^{c+1}$ determined in the preceding hierarchical classification step for $c = K-2$ and $K-3$ and the preceding non-hierarchical classification step for $c$ from $K-4$ to 2.
Based on the distances between $\widehat{\beta}^{c+1}_{i}$ and $\widehat{\beta}^{c+1}_{j}, 1 \leq i < j \leq c+1$, in the LSE $\widehat{\beta}^{c+1}$ of $\beta^{c+1}$, a pair of clusters, say $\big(G_{\ell}^{c+1}, G_{m}^{c+1} \big)$, with the minimal distance is merged into one cluster.
The clusters of robots $g^{c}$ are, thus, determined by the clusters $\{ G_{1}^{c}, \dots, G_{c}^{c} \}$ with $G_{1}^{c} = G_{\ell}^{c+1} \cup G_{m}^{c+1}$ and $ G_{k}^{c}$, $k = 2, \dots, c$, in one-to-one correspondence with the rest $G_{k}^{c+1}$'s.
For a WMPRC model $Y = X \beta^{c} + \varepsilon$ with $g^{c}$, $\beta^{c}$ is naturally estimated by the LSE $\widehat{\beta}^{c}$.
A full description of the proposed hierarchical classification is given in Algorithm \ref{alg:hier}.
\RestyleAlgo{ruled}
\begin{algorithm}
\caption{The hierarchical classification of robots}\label{alg:hier}
  \eIf{$c = K$}{
  $g^c \gets (1, \dots, c)^{\top}$\;
   Fit $Y = X \beta^{c} + \varepsilon$ and compute $\widehat{\beta}^{c}$, $\widehat{\textsc{mspe}}(c,g^{c})$ and $\widehat{\textsc{mspeb}}(c,g^{c})$\;
   }{$(\ell, m) \gets \argmin\limits_{i<j} \Big| \widehat{\beta}^{c+1}_{i} - \widehat{\beta}^{c+1}_{j} \Big|$ \;
   Determine $g^{c}$ by $\{ G^{c}_1, \dots, G^{c}_{c} \}$ with $G_{1}^{c} = G_{\ell}^{c+1} \cup G_{m}^{c+1}$ and $ G_{k}^{c}$, $k = 2, \dots, c$, in one-to-one correspondence with the rest $G_{k}^{c+1}$'s\;
   Fit $Y = X \beta^{c} + \varepsilon$ with $g^c$ and compute $\widehat{\beta}^{c}$, $\widehat{\textsc{mspe}}(c,g^{c})$ and $\widehat{\textsc{mspeb}}(c,g^{c})$ \;
   }{
  }
\end{algorithm}

To reduce the misclassification of robots in our hierarchical classification, a non-hierarchical classification is performed on WMPRC models $Y = X \beta^c + \varepsilon$ with $g^c$ determined in the hierarchical classification step for $c$ from $K-2$ to 2.
In this part, estimated robot strengths are retained the same for robots in $\{ G_{k}^c : |G_{k}^c| = 1, k = 1, \dots, c\}$.
The strength of robot $i$ in $\{ G_{k}^c : |G_{k}^c| \geq 2, k = 1, \dots, c\}$ is reestimated by the $i$th element of the LSE of $\beta^{c+1}$ in a WMPRC model $Y = X \beta^{c+1} + \varepsilon$ with $g^{c+1}$ set to $g^c$ and the $i$th element of $g^{c+1}$ replaced by $c+1$.
The original clusters of robots $g^{c}$ are further updated by performing the centroid linkage clustering method on these estimated robot strengths.
Thus, new estimated robot strengths can be directly obtained from the LSE of the regression coefficients in the resulting WMPRC model.
This process is repeated until updated clusters of robots are the same as those in the preceding iterations.
In practice, such a non-hierarchical classification greatly reduces the misclassification of robots when the number of clusters $c$ ($\geq c_o$) is relatively small compared to the number of matches $M$.
The details of our non-hierarchical classification are described in Algorithm \ref{alg:nonhier}.

\begin{algorithm}
\caption{The non-hierarchical classification of robots}\label{alg:nonhier}
   $r \gets 0, \widehat{\beta}^c_{(0)} \gets \widehat{\beta}^c$, $\epsilon \gets \text{tolerance value}$\;
   \Repeat{$\min\limits_{0 \leq r_o < {r}} \Big\| \widehat{\beta}^{c}_{({r})}$ $-$ $\widehat{\beta}^{c}_{(r_o)} \Big\| < \epsilon$ 
}{ $r \gets r+1$\;
   $\widetilde{\beta}_{(r)} \gets \widehat{\beta}^{c}_{(r-1)}$\;
   \For{$i \in \{ G_{k}^{c} : |G_{k}^{c}| \geq 2, k = 1, \dots, c \}$}{
    $g^{c+1} \gets g^{c}$\;
    $g^{c+1}_i \gets c+1$\;
    Replace the $i$th element of $\widetilde{\beta}_{(r)}$ by the $i$th element of the LSE of $\beta^{c+1}$ in $Y = X \beta^{c+1} + \varepsilon$ with $g^{c+1}$ \;
    }
   Update $g^{c}$ by performing the centroid linkage clustering method on $\widetilde{\beta}_{(r)}$ and compute the LSE $\widehat{\beta}^{c}_{(r)}$ of $\beta^{c}$ in $Y = X \beta^{c} + \varepsilon$ with $g^{c}$\;
   }
   Fit $Y = X \beta^{c} + \varepsilon$ with $g^{c}$ determined by $\widehat{\beta}^{c}_{({r})}$ (the LSE $\widehat{\beta}^{c}$ of ${\beta}^{c}$)\;
\end{algorithm}

\begin{rem}
Similar to the mean-regression-based method of \cite{lin2012estimation}, any one of the MSPE estimates in (\ref{eq:mspey_hat})  can be used as an assessment for merging two clusters and reassigning robots to other clusters.
However, there may be more than one WMPRC model with the same minimal MSPE estimate for some numbers of clusters.
Furthermore, more computational time is needed to obtain WMPRC candidate models.
As shown in the simulation study, all the MSPE measures are also underestimated by their cross-validation estimates in the format of qualification stage data from the 2018 and 2019 FRC Houston and Detroit championships.
Therefore, the MSPE estimates are inappropriate to serve as assessments for the classification of robots.  \qed
\end{rem}

By performing the proposed classification system, we can determine WMPRC candidate models $Y = X \beta^c + \varepsilon$ with $g^c$, $c = 2, \dots, K$.
Same as the model selection criteria in the LCT method, the parameteres $c_o$, $g^o$ and $\beta^o$ in model (\ref{eq:lnmodel}) are further estimated by $\widehat{c} = \arg\min_c \widehat{\textsc{mspe}}(c,g^c)$, ${g}^{\widehat{c}}$ and the LSE $\widehat{\beta}^{\widehat{c}}$, respectively, where $\widehat{\textsc{mspe}}(c,g^c)$ is one of the estimated functions in (\ref{eq:mspey_hat}).
Since these MSPE criteria tend to overestimate $c_o$, their alternatives are established based on the following estimated MSPEB functions:
\begin{align}
\widehat{\textsc{mspeb}}_{\textsc{y}}\big(c,g^c\big)  = & \ln\left(\widehat{\textsc{mspe}}_{\textsc{y}}\big(c,g^c\big)\right) + \frac{c \ln (M)}{M},\label{eq:mspeby_hat} \\
\widehat{\textsc{mspeb}}_{\textsc{p}}\big(c,g^c\big)  =&  \ln\left(\widehat{\textsc{mspe}}_{\textsc{p}}\big(c,g^c\big)\right) + \frac{c \ln (M)}{M},\text{ and} \nonumber \\
\widehat{\textsc{mspeb}}_{\textsc{d}}\big(c,g^c\big)  = & \ln\left(\widehat{\textsc{mspe}}_{\textsc{d}}\big(c,g^c\big)\right) + \frac{c \ln (M)}{M}, c = 2, \dots, K. \nonumber
\end{align}
As in the Schwarz criterion \citep{S1978} for linear models with normal errors, the basic idea behind such criteria is to introduce penalties for overparameterized models.
Thus, we can estimate the parameters $c_o$, $g^o$ and $\beta^o$ with $\widetilde{c} = \arg\min_c \widehat{\textsc{mspeb}}(c,g^c)$, ${g}^{\widetilde{c}}$ and the LSE $\widehat{\beta}^{\widetilde{c}}$, respectively, where $\widehat{\textsc{mspeb}}(c,g^c)$ is one of the estimated functions in (\ref{eq:mspeby_hat}).
In our data analysis and simulation study, all estimated MSPE and MSPEB functions exhibit U-shaped patterns over $c$.
The established criteria are also found to be useful for selecting optimal models from WMPRC candidate models.
\end{subsection}

\begin{subsection}{Estimation Procedure}
To enhance performance of the LCT method, we develop a more effective method.
The first part of this method is to perform a classification system for the determination of candidate models.
Another part is to establish the MSPE and MSPEB criteria for the selection of optimal models from these candidate models.

To avoid verbosity, the subscripts $\text{Y}$, $\text{P}$ and $\text{D}$ in the MSPE and MSPEB functions (including their estimates) are dropped in the following algorithm of the proposed estimation procedure:
\RestyleAlgo{ruled}

\begin{algorithm}
\caption{Algorithm of the proposed estimation procedure}\label{alg:tcl}
\Begin{
 \For{$c \gets K$ \KwTo $2$}{
  \eIf{$c = K$}{
  $g^c \gets (1, \dots, c)^{\top}$\;
   Fit $Y = X \beta^{c} + \varepsilon$ and compute $\widehat{\beta}^{c}$, $\widehat{\textsc{mspe}}(c,g^{c})$ and $\widehat{\textsc{mspeb}}(c,g^{c})$\;
   }{$(\ell, m) \gets \argmin\limits_{i<j} \Big| \widehat{\beta}^{c+1}_{i} - \widehat{\beta}^{c+1}_{j} \Big|$ \;
   Determine $g^{c}$ by $\{ G^{c}_1, \dots, G^{c}_{c} \}$ with $G_{1}^{c} = G_{\ell}^{c+1} \cup G_{m}^{c+1}$ and $ G_{k}^{c}$, $k = 2, \dots,$ $c$, in one-to-one correspondence with the rest $G_{k}^{c+1}$'s\;
   Fit $Y = X \beta^{c} + \varepsilon$ with $g^c$ and compute $\widehat{\beta}^{c}$, $\widehat{\textsc{mspe}}(c,g^{c})$ and $\widehat{\textsc{mspeb}}(c,g^{c})$ \;
   \If{$c \leq K-2$}{
   $r \gets 0, \widehat{\beta}^c_{(0)} \gets \widehat{\beta}^c$, $\epsilon \gets \text{tolerance value}$\;
   \Repeat{$\min\limits_{0 \leq r_o < {r}} \Big\| \widehat{\beta}^{c}_{({r})}$ $-$ $\widehat{\beta}^{c}_{(r_o)} \Big\| < \epsilon$
}{ $r \gets r+1$\;
   $\widetilde{\beta}_{(r)} \gets \widehat{\beta}^{c}_{(r-1)}$\;
   \For{$i \in \{ G_{k}^{c} : |G_{k}^{c}| \geq 2, k = 1, \dots, c \}$}{
    $g^{c+1} \gets g^{c}$\;
    $g^{c+1}_i \gets c+1$\;
    Replace the $i$th element of $\widetilde{\beta}_{(r)}$ by the $i$th element of the LSE of $\beta^{c+1}$ in $Y = X \beta^{c+1} + \varepsilon$ with $g^{c+1}$ \;
    }
   Update $g^{c}$ by performing the centroid linkage clustering method on $\widetilde{\beta}_{(r)}$ and compute the LSE $\widehat{\beta}^{c}_{(r)}$ of $\beta^{c}$ in $Y = X \beta^{c} + \varepsilon$ with $g^{c}$\;
   }
   Fit $Y = X \beta^{c} + \varepsilon$ with $g^{c}$ determined by $\widehat{\beta}^{c}_{({r})}$ (the LSE $\widehat{\beta}^{c}$ of ${\beta}^{c}$) and compute $\widehat{\textsc{mspe}}(c,g^{c})$ and $\widehat{\textsc{mspeb}}(c,g^{c})$\;
   }
   }
 }
  $\widehat{c} \gets \arg\min_c \widehat{\textsc{mspe}}(c,g^c)$, $\widetilde{c} \gets \arg\min_c \widehat{\textsc{mspeb}}(c,g^c)$\;
  Estimate $(c_o, g^o,\beta^o)$ by $\left(\widehat{c}, {g}^{\widehat{c}}, \widehat{\beta}^{\widehat{c}}\right)$ or $\left(\widetilde{c}, {g}^{\widetilde{c}}, \widehat{\beta}^{\widetilde{c}}\right)$\;
  }
\end{algorithm}
\noindent The classification system in Algorithm \ref{alg:tcl} has been described in Algorithms \ref{alg:hier} and \ref{alg:nonhier}.

Different from the proposed hierarchical classification, the clusters of robots $g^c$ in the LCT method are determined by estimated robot strengths in the preceding hierarchical classification step for $c$ from $K-1$ to 2.
By incorporating our non-hierarchical classification into the LCT method, an alternative method can also be used to estimate the parameters $c_o$, $g^o$ and $\beta^o$.
This estimation procedure is displayed in Algorithm \ref{alg:alter} for clarity.
The results are not presented because its performance is generally inferior to that of the developed method in the simulation study.

\begin{algorithm}
\caption{Algorithm of an alternative estimation procedure}\label{alg:alter}
\Begin{
 \For{$c \gets K$ \KwTo $2$}{
  \eIf{$c = K$}{
  $g^c \gets (1, \dots, c)^{\top}$\;
   Fit $Y = X \beta^{c} + \varepsilon$ and compute $\widehat{\beta}^{c}$, $\widehat{\textsc{mspe}}(c,g^{c})$ and $\widehat{\textsc{mspeb}}(c,g^{c})$\;
   }{$(\ell, m) \gets \argmin\limits_{i<j} \Big| \widehat{\beta}^{c+1}_{i} - \widehat{\beta}^{c+1}_{j} \Big|$ \;
   Determine $g^{c}$ by $\{ G^{c}_1, \dots, G^{c}_{c} \}$ with $G_{1}^{c} = G_{\ell}^{c+1} \cup G_{m}^{c+1}$ and $ G_{k}^{c}$, $k = 2, \dots, c$, in one-to-one correspondence with the rest $G_{k}^{c+1}$'s\;
   Fit $Y = X \beta^{c} + \varepsilon$ with $g^c$ and compute $\widehat{\beta}^{c}$, $\widehat{\textsc{mspe}}(c,g^{c})$ and $\widehat{\textsc{mspeb}}(c,g^{c})$ \;
   }{
  }
 }
\For{$c \gets K-2$ \KwTo $2$}{
   $r \gets 0, \widehat{\beta}^c_{(0)} \gets \widehat{\beta}^c$,  $\epsilon \gets \text{tolerance value}$\;
   \Repeat{$\min\limits_{0 \leq r_o < {r}} \Big\| \widehat{\beta}^{c}_{({r})}$ $-$ $\widehat{\beta}^{c}_{(r_o)} \Big\| < \epsilon$ 
}{ $r \gets r+1$\;
   $\widetilde{\beta}_{(r)} \gets \widehat{\beta}^{c}_{(r-1)}$\;
   \For{$i \in \{ G_{k}^{c} : |G_{k}^{c}| \geq 2, k = 1, \dots, c \}$}{
    $g^{c+1} \gets g^{c}$\;
    $g^{c+1}_i \gets c+1$\;
    Replace the $i$th element of $\widetilde{\beta}_{(r)}$ by the $i$th element of the LSE of $\beta^{c+1}$ in $Y = X \beta^{c+1} + \varepsilon$ with $g^{c+1}$ \;
    }
   Update $g^{c}$ by performing the centroid linkage clustering method on $\widetilde{\beta}_{(r)}$ and compute the LSE $\widehat{\beta}^{c}_{(r)}$ of $\beta^{c}$ in $Y = X \beta^{c} + \varepsilon$ with $g^{c}$\;
   }
   Fit $Y = X \beta^{c} + \varepsilon$ with $g^{c}$ determined by $\widehat{\beta}^{c}_{({r})}$ (the LSE $\widehat{\beta}^{c}$ of ${\beta}^{c}$) and compute $\widehat{\textsc{mspe}}(c,g^{c})$ and $\widehat{\textsc{mspeb}}(c,g^{c})$\;
   }
    $\widehat{c} \gets \arg\min_c \widehat{\textsc{mspe}}(c,g^c)$, $\widetilde{c} \gets \arg\min_c \widehat{\textsc{mspeb}}(c,g^c)$\;
  Estimate $(c_o, g^o,\beta^o)$ by $\left(\widehat{c}, {g}^{\widehat{c}}, \widehat{\beta}^{\widehat{c}}\right)$ or $\left(\widetilde{c}, {g}^{\widetilde{c}}, \widehat{\beta}^{\widetilde{c}}\right)$\;
  }
\end{algorithm}

\begin{rem} \label{rem:fusion}
In the spirit of a pairwise fusion penalized least squares method of \cite{bondell2008simultaneous}, the robot strengths $\beta^{o}$ in model (\ref{eq:lnmodel}) can also be estimated by a minimizer of the following objective function with the constraint $\sum^{K}_{i=1}\beta_{i}=0$:
\begin{align}
\frac{1}{2}(Y-X\beta)^{\top}(Y-X\beta)+\mathop{\sum\sum}_{1\leq i<j\leq K}p(|\beta_{i}-\beta_{j}|,\lambda),\label{cpfusion}
\end{align}
where $\lambda$ is a non-negative tuning parameter and $p(\cdot,\lambda)$ is a penalty function such as the smoothly clipped absolute deviation of \cite{fan2001variable} and the minimax concave penalty of \cite{zhang2010nearly}. 
In implementation, this minimization is achieved by the alternating direction method of multipliers of \cite{boyd2011distributed} and the penalty parameter is chosen by the traditional BIC of \cite{S1978}. 
A model selected by the best criterion in the developed method outperforms its estimated model in the data analysis (see Table \ref{tab:fusion}), albeit an estimator obtained from the criterion in (\ref{cpfusion}) enjoys the oracle properties under some regularity conditions.
The conducted simulation also shows that our method provides better estimation of parameters and more accurate prediction of match scores (or outcomes) in the format of qualification stage data (see Table \ref{tab:fusion_simu}). 
The data analysis and simulation results of this pairwise fusion method are displayed in the supplementary material.\qed
\end{rem}

\end{subsection}

\begin{subsection}{Matching Index of the Nested Relation and Rank Correlation}
To measure the nested relation between clusters of the WMPRC models $Y = X \beta^c + \varepsilon$ with $g^c$ and $Y = X \beta^d + \varepsilon$ with $g^d$, a matching index of the nested relation (MINR) is proposed as follows:
\begin{align}
\widehat{\textsc{minr}}\big(g^{c},g^{d}\big) = & \frac{1}{K}  \sum_{i=1}^{K} \bigg( I\big(\widehat{g}_{i}^c = g_{i}^d \big) I(c > d) + I\big(g_{i}^c = \widehat{g}_{i}^d\big) I(c < d) \label{eq:MINR}   \\
&+   \frac{I\big(g_{i}^c = \widehat{g}_{i}^d \big) + I\big(\widehat{g}_{i}^c = g_{i}^d\big)}{2} I(c = d)\bigg) , \nonumber
\end{align}
where $\widehat{g}_{i}^c = \ell$ if $\left|\widehat{\beta}^c_{g_{i}^c} - \widehat{\beta}^d_{\ell} \right| = \min\limits_{1\leq j \leq d} \left|\widehat{\beta}^c_{g_{i}^c} - \widehat{\beta}^d_{j} \right|$ and $\widehat{g}_{i}^d = m$ if $\Big|\widehat{\beta}^d_{g_{i}^d} - \widehat{\beta}^c_{m} \Big| = \min\limits_{1\leq j \leq c} \left|\widehat{\beta}^d_{g_{i}^d} - \widehat{\beta}^c_{j} \right|$, $i = 1, \dots, K$, $\ell = 1, \dots, d$, $m = 1, \dots, c$.
Given the information of $c$ and $d$, one of the models with a smaller number of clusters is treated as a reference model.
The MINR investigates whether another model is in a more general form.
When both models have the same number of clusters, they serve as a reference model of each other.
In our simulation study, the true WMPRC model is treated as a reference model.
The main purpose is to introduce penalties for underparameterized models.
With this consideration, the MINR is modified as 
\begin{align}
\widehat{\textsc{minr}}\big(g^{c},g^o\big) = \frac{1}{K} \sum_{i=1}^{K} I\big(\widehat{g}_{i}^c = g_{i}^o\big), \label{eq:MINR_true}
\end{align}
where $\widehat{g}_{i}^c = \ell$ if $\left|\widehat{\beta}^c_{g_{i}^c} - \beta^o_{\ell} \right| = \min\limits_{1\leq j \leq c_o} \left|\widehat{\beta}^c_{g_{i}^c} - \beta^o_{j} \right|$, $i = 1, \dots, K$, $\ell = 1, \dots, c_o$.

In application, the rank correlation (RC) of \cite{H1987} has been used by \cite{LCT2021} to measure the monotonic association between robot strengths from any two models.
Since robot strengths are related to their intrinsic ranks, it is inappropriate to take a weight of 0.5 for those robots assigned to the same clusters.
For the considered ordinal scale measurements, we modify their rank correlation estimate of ${\beta}^{c}$ and ${\beta}^{d}$ as follows:
\begin{align}
\widehat{\textsc{rc}}\big({\beta}^{c},{\beta}^{d} \big) = & \frac{1}{K(K-1)} \mathop{\sum\sum}_{i \neq j} \Big(I\Big( \Big(\widehat{\beta}^c_{{g}_{i}^c}- \widehat{\beta}^c_{{g}_{j}^c} \Big)  \Big(\widehat{\beta}^d_{{g}_{i}^d}- \widehat{\beta}^d_{{g}_{j}^d}\Big)  > 0 \Big)  \label{eq:RC}\\
& + I\Big( \widehat{\beta}^c_{{g}_{i}^c} = \widehat{\beta}^c_{{g}_{j}^c} \text{ and } \widehat{\beta}^d_{{g}_{i}^d} = \widehat{\beta}^d_{{g}_{j}^d}  \Big)  \Big). \nonumber
\end{align}
Naturally, a rank correlation estimate of $\beta^c$ and $\beta^o$ can be computed as in (\ref{eq:RC}) with $\widehat{\beta}^d_{g^d_i}$'s replaced by $\beta^o_{g^o_i}$'s.
\begin{rem}
In principle, the rank-based index is not designed to measure the nested relation between clusters from any two models.
As shown in Figures \ref{fig:MINR} and \ref{fig:RC}, the strength of monotonic association is not necessarily implied by the strength of nested relation and vice versa.\qed
\end{rem}
\end{subsection}
\end{section}

\begin{section}{Data Analysis and Simulation Study} \label{sec:analysis}
We applied the developed method (hereinafter referred to as the TCL method) and the LCT method of \cite{LCT2021} to qualification stage and playoff stage data from the 2018 and 2019 FRC Houston and Detroit championships.
A Monte Carlo simulation was further conducted to investigate the performance of both the methods in identifying clusters of robot strengths and predicting an outcome, outcome probabilities and a difference in scores.

\begin{subsection}{Application}
In the divisions of the 2018 and 2019 FRC Houston and Detroit championships, there were 110 to 114 scheduled matches involving 66 to 68 robots in the qualification stage.
Under the playoff system of the FRC, 14 to 18 matches were played by top eight robotics teams and their self-selected alliance partners.
Details of the data description and processing can be found in \cite{LCT2021}.
In the data analysis, one match with extremely high scores in the Turing division of the 2018 FRC Houston championship was removed to ensure better quality of data.
We further computed the MSPE and MSPEB estimates of each WMPRC candidate model as in (\ref{eq:mspey_hat}) and (\ref{eq:mspeby_hat}) and the MINR and RC of two selected optimal models as in (\ref{eq:MINR}) and (\ref{eq:RC}).

In Figures \ref{plot:2018_H} -- \ref{plot:2019_D}, it can be seen that the estimated MSPE and MSPEB curves are slightly serrated convex functions of the number of clusters $c$.
Except for the $\text{MSPE}_{\text{D}}$ criterion, the rest criteria have unique minimizers.
For those WMPRC candidate models with the minimal $\text{MSPE}_{\text{D}}$ estimate, a model with the smallest number of clusters is selected in implementation.
The parameters of selected models may be very different, albeit any one of estimated MSPE curves in the LCT method is close to the corresponding one in the TCL method.
With the same MSPE criterion for model selection, Tables \ref{tab:WMPRC2-MD} -- \ref{tab:HnH-MY} show that the averages of estimated numbers of clusters, which were computed over the divisions in the same tournament, are comparable in both the LCT and TCL methods.
In the TCL method, the corresponding MSPEB criteria of the MSPE criteria are further found to have noticeably smaller averages of estimated numbers of clusters.
The MSPE estimates of a model selected by any one of the MSPE criteria in the LCT method are comparable to the corresponding estimates of a model selected by the same criterion in the TCL method.
Moreover, models selected by the $\text{MSPE}_{\text{D}}$ ($\text{or PCP}$), $\text{MSPE}_{\text{P}}$ and $\text{MSPE}_{\text{Y}}$ criteria have the minimal $\text{MSPE}_{\text{D}}$ ($\text{or maximal PCP}$), $\text{MSPE}_{\text{P}}$ and $\text{MSPE}_{\text{Y}}$ estimates, respectively.
It is noted that the MSPE estimates of the WMPR model (see Table \ref{tab:WMPR}) are much larger than the corresponding estimates of selected optimal models in both the LCT and TCL methods.
\begin{figure}[htbp]
\centering
 \includegraphics[width=14cm,height=9.5cm]{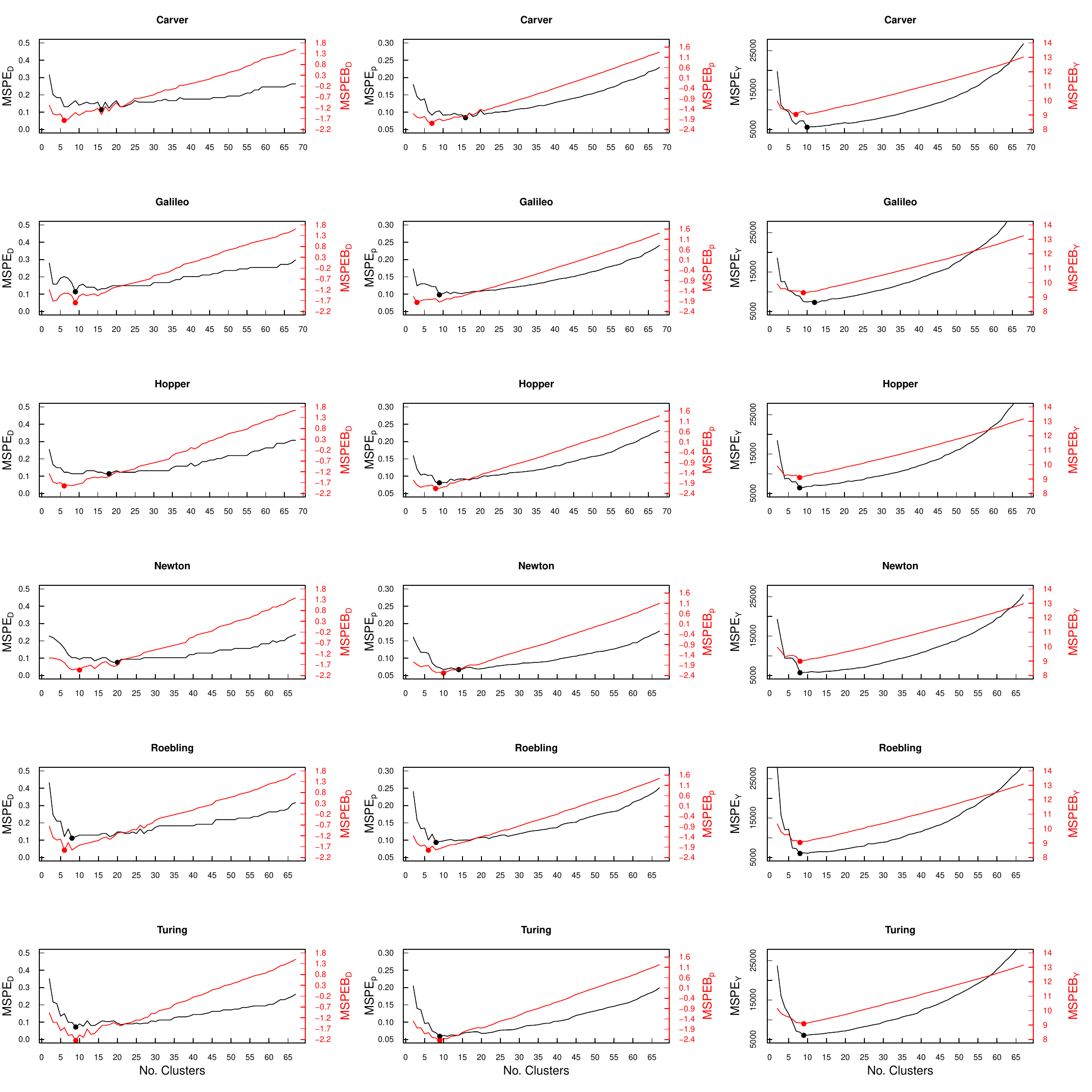}
 \caption{{\footnotesize The estimated MSPE curves (black line) and MSPEB curves (red line), accompanied with their minima (black and red dots), in the TCL method for the divisions of the 2018 FRC Houston championship.}} \label{plot:2018_H} 
\end{figure}

\begin{figure}[htbp]
\centering
 \includegraphics[width=14cm,height=9.5cm]{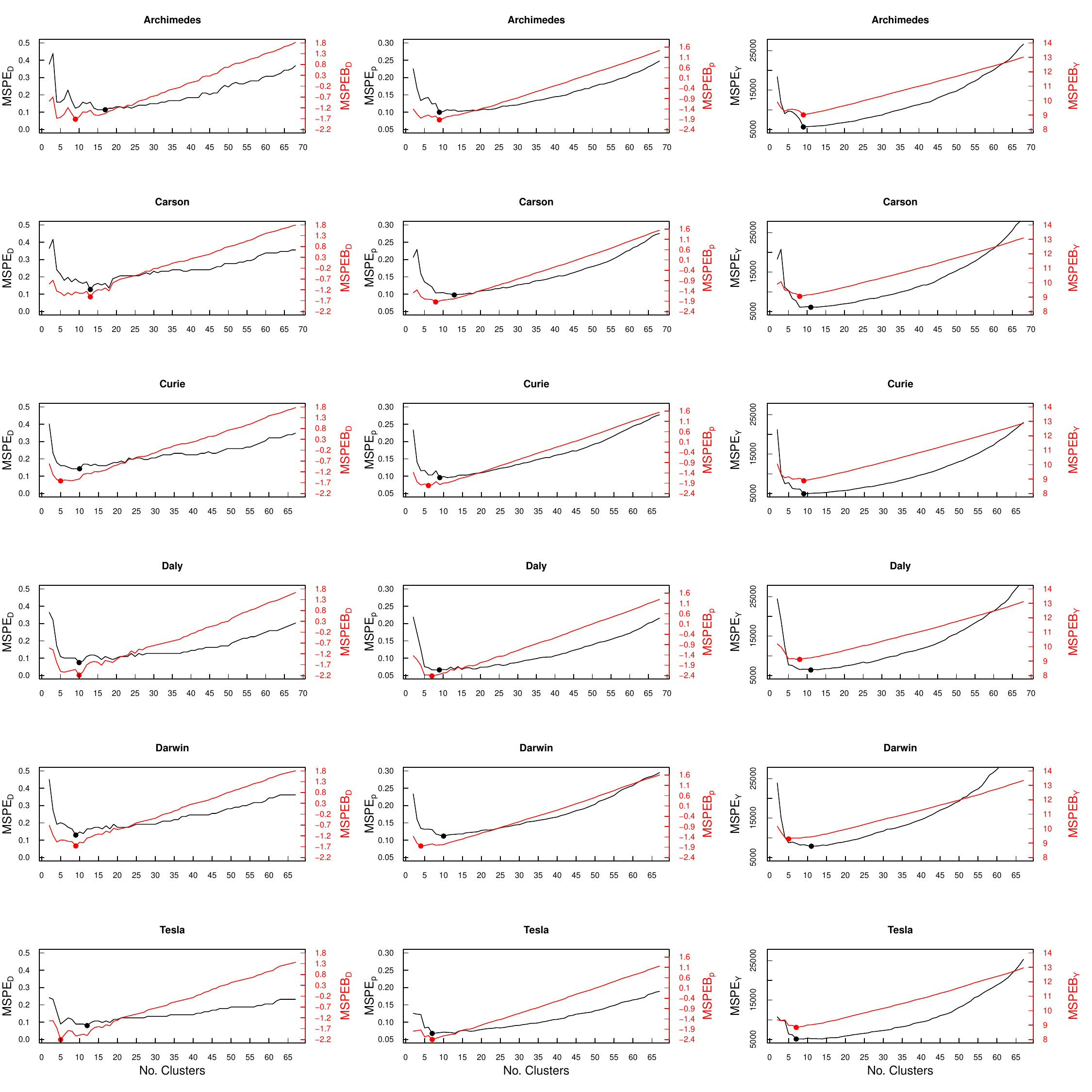}
 \caption{{\footnotesize The estimated MSPE curves (black line) and MSPEB curves (red line), accompanied with their minima (black and red dots), in the TCL method for the divisions of the 2018 FRC Detroit championship.}} \label{plot:2018_D} 
\end{figure}

\begin{figure}[htbp]
\centering
 \includegraphics[width=14cm,height=9.5cm]{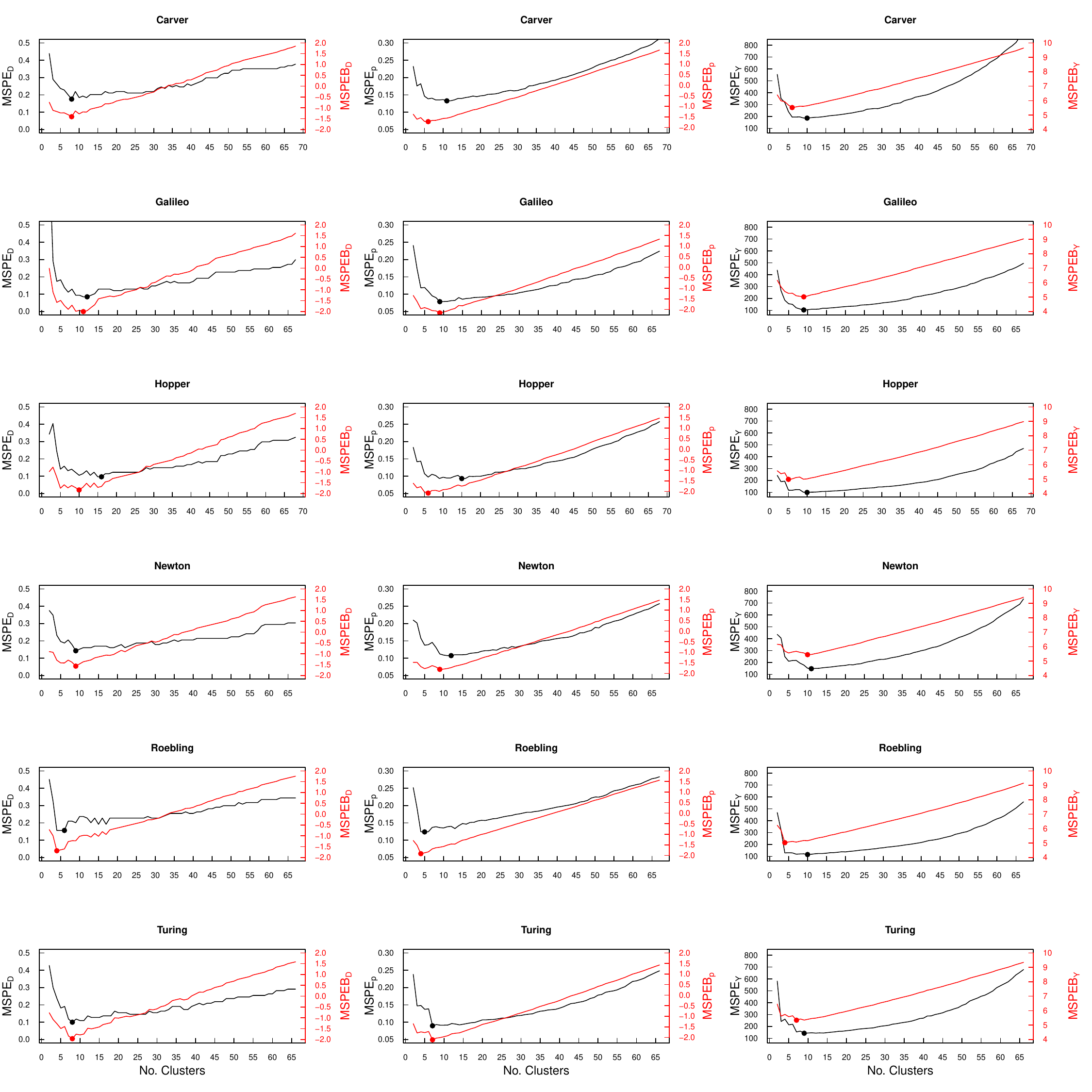}
 \caption{{\footnotesize The estimated MSPE curves (black line) and MSPEB curves (red line), accompanied with their minima (black and red dots), in the TCL method for the divisions of the 2019 FRC Houston championship.}} \label{plot:2019_H} 
\end{figure}

\begin{figure}[htp]
\centering
 \includegraphics[width=14cm,height=9.5cm]{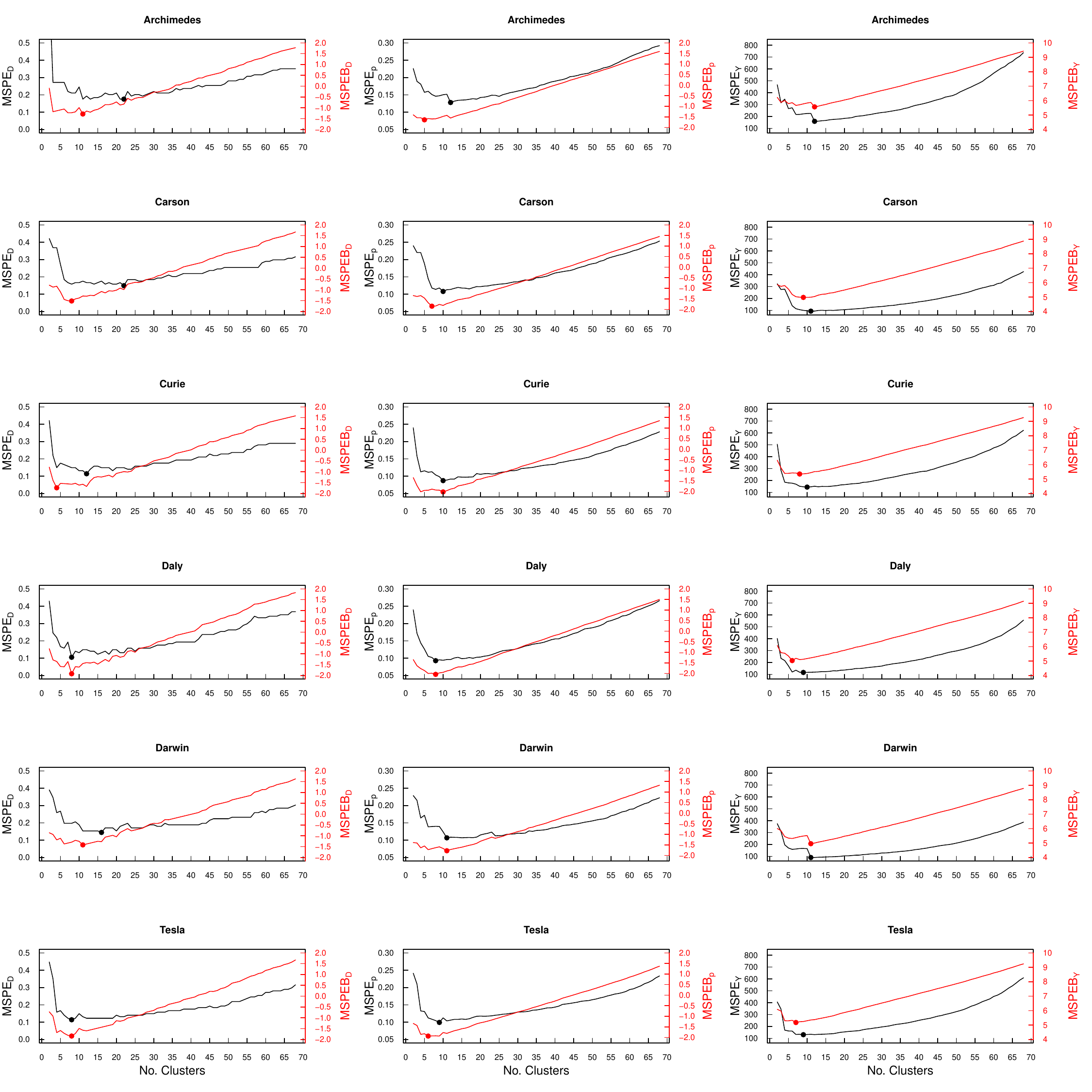}
 \caption{{\footnotesize The estimated MSPE curves (black line) and MSPEB curves (red line), accompanied with their minima (black and red dots), in the TCL method for the divisions of the 2019 FRC Detroit championship.}} \label{plot:2019_D} 
\end{figure}

\begin{table}[htbp]
  \tiny
  \centering
  \caption{The numbers of clusters in models selected by the $\text{MSPE}_{\text{D}}$ criterion in the LCT method, the $\text{PCP}$, $\text{MSPE}_{\text{P}}$ and $\text{MSPE}_{\text{Y}}$ estimates of these selected optimal models and the numbers of correct predictions of semifinalists, finalists and champions in the playoffs for the 2018 and 2019 FRC championships.}
    \resizebox{!}{3.73cm}{%
    \begin{tabular}{lllrrrrrrrr}
    \toprule    
     Season & Tournament & Division & $\widehat{c}$ & \multicolumn{1}{c}{$\widehat{\textsc{pcp}}$} & \multicolumn{1}{c}{$\widehat{\textsc{mspe}}_{\textsc{p}}$} & \multicolumn{1}{c}{$\widehat{\textsc{mspe}}_{\textsc{y}}$} & \multicolumn{1}{c}{Semifinalists}  &\multicolumn{1}{c}{Finalists} & \multicolumn{1}{c}{Champions}  \\
\midrule
    2018  & Houston & Carver & 12    & 87.7\% & 0.083 & 5681.5  & 3         & 0     & 1      \\
          &       & Galileo & 8     & 87.7\% & 0.101 & 7068.9  & 2     & 2        & 1 \\
          &       & Hopper & 21    & 88.6\% & 0.099 & 7924.3  & 3        & 2         & 0 \\
          &       & Newton & 14    & 92.4\% & 0.070 & 5979.6 & 3        & 2          & 0 \\
          &       & Roebling & 11    & 90.6\% & 0.090 & 6105.8  & 2         & 1          & 0 \\
          &       & Turing & 12    & 92.8\% & 0.061 & 6285.1  & 2         & 1        & 0 \\
          \cline{2-10}
          & Detroit & Archimedes & 6     & 88.6\% & 0.102 & 6124.8  & 4        & 0          & 1 \\
          &       & Carson & 12    & 86.4\% & 0.098 & 6027.4  & 4        & 0       & 1    \\
          &       & Curie & 6     & 87.5\% & 0.090 & 6081.9  & 2        & 1        & 1 \\
          &       & Daly  & 16    & 90.8\% & 0.069 & 6740.3  & 2         & 2         & 0     \\
          &       & Darwin & 8     & 85.3\% & 0.118 & 7499.7  & 0        & 1         & 0 \\
          &       & Tesla & 7     & 92.9\% & 0.073 & 5991.3  & 3          & 1         & 0 \\
             \midrule
    Average &       &       & 11.1  & 89.3\% & 0.088 & 6459.2  & 2.5      & 1.1     & 0.4    \\
        \midrule
        \midrule
    2019  & Houston & Carver & 12    & 82.5\% & 0.136 & 191.8  & 4              & 1     & 1 \\
          &       & Galileo & 9     & 91.5\% & 0.082 & 103.3 & 2     & 2          & 0 \\
          &       & Hopper & 12    & 90.4\% & 0.095 & 102.2  & 4         & 1     & 1 \\
          &       & Newton & 8     & 85.7\% & 0.109 & 149.0  & 3        & 1     & 1 \\
          &       & Roebling & 4     & 84.4\% & 0.121 & 131.7  & 3        & 2        & 0 \\
          &       & Turing & 7     & 89.1\% & 0.091 & 141.2  & 2        & 1     & 1 \\
          \cline{2-10}
          & Detroit & Archimedes & 12    & 84.2\% & 0.133 & 165.4  & 1          & 1        & 1 \\
          &       & Carson & 9     & 86.0\% & 0.103 & 88.8   & 2     & 1        & 0 \\
          &       & Curie & 14    & 86.0\% & 0.099 & 145.0  & 4         & 0         & 1     \\
          &       & Daly  & 11    & 86.8\% & 0.092 & 117.0  & 0          & 0          & 0 \\
          &       & Darwin & 10    & 86.4\% & 0.104 & 90.5    & 2          & 2         & 1 \\
          &       & Tesla & 8     & 87.7\% & 0.106 & 132.2  & 2          & 1     & 1 \\
    \midrule
    Average &       &       & 9.7   & 86.7\% & 0.106 & 129.9  & 2.4     & 1.1    & 0.7    \\
     \bottomrule
    \end{tabular}%
    }
  \label{tab:WMPRC2-MD}%
\end{table}%

\begin{table}[htbp]
  \tiny
  \centering
  \caption{The numbers of clusters in models selected by the $\text{MSPE}_{\text{P}}$ criterion in the LCT method, the $\text{PCP}$, $\text{MSPE}_{\text{P}}$ and $\text{MSPE}_{\text{Y}}$ estimates of these selected optimal models and the numbers of correct predictions of semifinalists, finalists and champions in the playoffs for the 2018 and 2019 FRC championships.}
        \resizebox{!}{3.73cm}{%
    \begin{tabular}{lllrrrrrrrr}
    \toprule    
    Season & Tournament & Division & $\widehat{c}$ & \multicolumn{1}{c}{$\widehat{\textsc{pcp}}$} & \multicolumn{1}{c}{$\widehat{\textsc{mspe}}_{\textsc{p}}$} & \multicolumn{1}{c}{$\widehat{\textsc{mspe}}_{\textsc{y}}$} & \multicolumn{1}{c}{Semifinalists}  &\multicolumn{1}{c}{Finalists} & \multicolumn{1}{c}{Champions}  \\
\midrule
    2018  & Houston & Carver & 12    & 87.7\% & 0.083 & 5681.5  & 3          & 0          & 1    \\
          &       & Galileo & 14    & 85.1\% & 0.099 & 7698.4 & 2          & 2          & 1      \\
          &       & Hopper & 10    & 86.8\% & 0.089 & 6634.4  & 3          & 2         & 0     \\
          &       & Newton & 16    & 91.5\% & 0.070 & 6163.3  & 3          & 2          & 0      \\
          &       & Roebling & 10    & 87.1\% & 0.089 & 6152.1  & 2         & 1         & 0      \\
          &       & Turing & 11    & 90.1\% & 0.058 & 6297.2 & 2         & 1         & 0      \\
          \cline{2-10}
          & Detroit & Archimedes & 9     & 86.8\% & 0.100 & 5716.8  & 4         & 0       & 1 \\
          &       & Carson & 13    & 85.5\% & 0.098 & 6102.3 & 4         & 0         & 1 \\
          &       & Curie & 7     & 83.9\% & 0.084 & 5771.9  & 3.5        & 1     & 1 \\
          &       & Daly  & 16    & 90.8\% & 0.069 & 6740.3 & 2     & 2       & 0     \\
          &       & Darwin & 11    & 85.3\% & 0.113 & 7597.9  & 0        & 1          & 0 \\
          &       & Tesla & 8     & 87.5\% & 0.066 & 5309.2  & 3        & 0     & 0 \\
          \midrule
    Average &       &       & 11.4  & 87.3\% & 0.085 & 6322.1  & 2.6      & 1.0      & 0.4    \\
        \midrule
        \midrule
    2019  & Houston & Carver & 11    & 81.6\% & 0.133 & 189.5  & 4            & 1     & 1 \\
          &       & Galileo & 8     & 90.6\% & 0.081 & 105.5  & 2     & 2        & 0 \\
          &       & Hopper & 13    & 90.4\% & 0.091 & 102.2  & 4          & 1     & 1 \\
          &       & Newton & 8     & 85.7\% & 0.109 & 149.0  & 3          & 1     & 1 \\
          &       & Roebling & 5     & 84.4\% & 0.119 & 131.3  & 3          & 2         & 0 \\
          &       & Turing & 9     & 87.3\% & 0.091 & 137.1  & 1        & 2          & 1 \\
          \cline{2-10}
          & Detroit & Archimedes & 7     & 83.3\% & 0.125 & 169.4  & 2         & 1     & 1 \\
          &       & Carson & 9     & 86.0\% & 0.103 & 88.8   & 2     & 1        & 0 \\
          &       & Curie & 12    & 84.2\% & 0.096 & 144.2  & 4        & 0      & 1 \\
          &       & Daly  & 11    & 86.8\% & 0.092 & 117.0  & 0     & 0     & 0      \\
          &       & Darwin & 5     & 82.0\% & 0.101 & 127.3  & 1         & 2        & 1 \\
          &       & Tesla & 13    & 87.7\% & 0.102 & 133.4  & 2         & 1     & 1 \\
    \midrule
    Average &       &       & 9.3   & 85.8\% & 0.104 & 132.9  & 2.3   & 1.2    & 0.7    \\
     \bottomrule
    \end{tabular}%
    }
  \label{tab:WMPRC2-MP}%
\end{table}%

\begin{table}[htbp]
  \tiny
  \centering
  \caption{The numbers of clusters in models selected by the $\text{MSPE}_{\text{Y}}$ criterion in the LCT method, the $\text{PCP}$, $\text{MSPE}_{\text{P}}$ and $\text{MSPE}_{\text{Y}}$ estimates of these selected optimal models and the numbers of correct predictions of semifinalists, finalists and champions in the playoffs for the 2018 and 2019 FRC championships.}
    \resizebox{!}{3.73cm}{%
    \begin{tabular}{lllrrrrrrrr}
    \toprule    
    Season & Tournament & Division & $\widehat{c}$ & \multicolumn{1}{c}{$\widehat{\textsc{pcp}}$} & \multicolumn{1}{c}{$\widehat{\textsc{mspe}}_{\textsc{p}}$} & \multicolumn{1}{c}{$\widehat{\textsc{mspe}}_{\textsc{y}}$} & \multicolumn{1}{c}{Semifinalists}  &\multicolumn{1}{c}{Finalists} & \multicolumn{1}{c}{Champions}  \\
\midrule
    2018  & Houston & Carver & 11    & 85.1\% & 0.091 & 5641.4 & 3         & 0        & 1 \\
          &       & Galileo & 8     & 87.7\% & 0.101 & 7068.9 & 2     & 2          & 1 \\
          &       & Hopper & 10    & 86.8\% & 0.089 & 6634.4 & 3          & 2        & 0 \\
          &       & Newton & 10    & 87.9\% & 0.078 & 5780.7  & 3        & 2      & 1 \\
          &       & Roebling & 11    & 90.6\% & 0.090 & 6105.8 & 2         & 1        & 0 \\
          &       & Turing & 12    & 92.8\% & 0.061 & 6285.1  & 2        & 1         & 0 \\
          \cline{2-10}
          & Detroit & Archimedes & 9     & 86.8\% & 0.100 & 5716.8  & 4        & 0        & 1 \\
          &       & Carson & 10    & 85.5\% & 0.101 & 5984.5  & 4        & 0       & 1 \\
          &       & Curie & 10    & 84.8\% & 0.092 & 4993.6  & 4        & 1         & 1 \\
          &       & Daly  & 11    & 88.2\% & 0.072 & 6479.8  & 2          & 2     & 0     \\
          &       & Darwin & 8     & 85.3\% & 0.118 & 7499.7  & 0         & 1          & 0 \\
          &       & Tesla & 12    & 91.1\% & 0.071 & 5117.4  & 2         & 1          & 0 \\
          \midrule
    Average &       &       & 10.2  & 87.7\% & 0.089 & 6109.0 & 2.6     & 1.1     & 0.5    \\
\midrule
\midrule
    2019  & Houston & Carver & 7     & 80.7\% & 0.137 & 185.5  & 4        & 1     & 1 \\
          &       & Galileo & 9     & 91.5\% & 0.082 & 103.3 & 2        & 2       & 0 \\
          &       & Hopper & 10    & 86.8\% & 0.100 & 100.2 & 4        & 1     & 1 \\
          &       & Newton & 11    & 83.9\% & 0.109 & 148.2  & 3          & 1     & 1 \\
          &       & Roebling & 9     & 78.1\% & 0.140 & 115.9  & 3         & 2         & 0 \\
          &       & Turing & 9     & 87.3\% & 0.091 & 137.1  & 1        & 2         & 1 \\
          \cline{2-10}
          & Detroit & Archimedes & 10    & 82.5\% & 0.131 & 160.1  & 2         & 1     & 1 \\
          &       & Carson & 9     & 86.0\% & 0.103 & 88.8   & 2     & 1       & 0 \\
          &       & Curie & 12    & 84.2\% & 0.096 & 144.2  & 4         & 0         & 1 \\
          &       & Daly  & 8     & 85.1\% & 0.097 & 114.4  & 0          & 1       & 0 \\
          &       & Darwin & 12    & 86.4\% & 0.105 & 89.1   & 1     & 2    & 1 \\
          &       & Tesla & 10    & 85.1\% & 0.104 & 128.4  & 2        & 1     & 1 \\
\midrule
    Average &       &       & 9.7   & 84.8\% & 0.108 & 126.3  & 2.3     & 1.3     & 0.7 \\
     \bottomrule
    \end{tabular}%
    }
  \label{tab:WMPRC2-MY}%
\end{table}%

\begin{table}[htbp]
  \tiny
  \centering
  \caption{The numbers of clusters in models selected by the (i) ${\text{MSPE}}_{\text{D}}$ and (ii) ${\text{MSPEB}}_{\text{D}}$  criteria in the TCL method, the $\text{PCP}$, $\text{MSPE}_{\text{P}}$ and $\text{MSPE}_{\text{Y}}$ estimates of these selected optimal models and the numbers of correct predictions of semifinalists, finalists and champions in the playoffs for the 2018 and 2019 FRC championships.}
    \resizebox{\columnwidth}{!}{%
    \begin{tabular}{lllrrrrrrrrrrrrrr}
    \toprule
  &  &  &  (i) & & &  &  (ii) & & & & (i) & (ii) & (i) & (ii) & (i) & (ii)  \\
    Season & Tournament & Division & $\widehat{c}$ & \multicolumn{1}{c}{$\widehat{\textsc{pcp}}$} & \multicolumn{1}{c}{$\widehat{\textsc{mspe}}_{\textsc{p}}$} & \multicolumn{1}{c}{$\widehat{\textsc{mspe}}_{\textsc{y}}$} & $\widetilde{c}$ & \multicolumn{1}{c}{$\widehat{\textsc{pcp}}$} & \multicolumn{1}{c}{$\widehat{\textsc{mspe}}_{\textsc{p}}$} & \multicolumn{1}{c}{$\widehat{\textsc{mspe}}_{\textsc{y}}$} & \multicolumn{2}{l}{Semifinalists}  &\multicolumn{2}{l}{Finalists} & \multicolumn{2}{l}{Champions}  \\
\midrule
    2018  & Houston & Carver & 16    & 88.6\% & 0.084 & 6087.9 & 6     & 86.8\% & 0.102 & 7191.7 & 3     & 3     & 0     & 0     & 1     & 1 \\
          &       & Galileo & 9     & 88.6\% & 0.098 & 7541.4 & 9     & 88.6\% & 0.098 & 7541.4 & 2     & 2     & 2     & 2     & 1     & 1 \\
          &       & Hopper & 8     & 88.6\% & 0.083 & 6447.7 & 6     & 87.7\% & 0.102 & 7933.3 & 3     & 3     & 2     & 2     & 0     & 0 \\
          &       & Newton & 19    & 92.4\% & 0.069 & 6462.7 & 10    & 90.6\% & 0.067 & 5803.8 & 3     & 3     & 2     & 2     & 1     & 0 \\
          &       & Roebling & 8     & 88.8\% & 0.094 & 6027.1 & 6     & 87.9\% & 0.100 & 7393.4 & 2     & 2     & 1     & 1     & 0     & 1 \\
          &       & Turing & 9     & 92.8\% & 0.059 & 6065.3 & 9     & 92.8\% & 0.059 & 6065.3 & 1     & 1     & 1     & 1     & 0     & 0 \\
          \cline{2-17}
          & Detroit & Archimedes & 15    & 88.6\% & 0.103 & 6051.1 & 9     & 87.7\% & 0.100 & 5646.8 & 4     & 4     & 0     & 0     & 1     & 1 \\
          &       & Carson & 13    & 87.3\% & 0.098 & 6125.0 & 13    & 87.3\% & 0.098 & 6125.0 & 4     & 4     & 0     & 0     & 1     & 1 \\
          &       & Curie & 8     & 85.7\% & 0.116 & 6119.7 & 5     & 83.9\% & 0.117 & 7767.1 & 2     & 2     & 1     & 2     & 1     & 1 \\
          &       & Daly  & 10    & 92.5\% & 0.068 & 6598.1 & 10    & 92.5\% & 0.068 & 6598.1 & 3     & 3     & 2     & 2     & 0     & 0 \\
          &       & Darwin & 9     & 87.1\% & 0.115 & 8263.4 & 9     & 87.1\% & 0.115 & 8263.4 & 0     & 0     & 1     & 1     & 0     & 0 \\
          &       & Tesla & 12    & 92.0\% & 0.070 & 5208.3 & 5     & 91.1\% & 0.083 & 6383.9 & 2     & 3     & 0     & 2     & 0     & 1 \\
          \midrule
    Average &       &       & 11.3  & 89.4\% & 0.088 & 6416.5 & 8.1   & 88.7\% & 0.092 & 6892.8 & 2.4   & 2.5   & 1.0   & 1.3   & 0.5   & 0.6 \\
\midrule
\midrule
    2019  & Houston & Carver & 8     & 82.5\% & 0.136 & 196.5 & 8     & 82.5\% & 0.136 & 196.5 & 4     & 4     & 0     & 0     & 1     & 1 \\
          &       & Galileo & 11    & 91.5\% & 0.079 & 109.1 & 11    & 91.5\% & 0.079 & 109.1 & 2     & 2     & 2     & 2     & 0     & 0 \\
          &       & Hopper & 15    & 90.4\% & 0.093 & 108.0 & 10    & 89.5\% & 0.097 & 99.3  & 4     & 4     & 1     & 0     & 1     & 1 \\
          &       & Newton & 9     & 85.7\% & 0.112 & 178.7 & 9     & 85.7\% & 0.112 & 178.7 & 3     & 3     & 1     & 1     & 1     & 1 \\
          &       & Roebling & 4     & 84.4\% & 0.124 & 129.0 & 4     & 84.4\% & 0.124 & 129.0 & 3     & 3     & 2     & 2     & 0     & 0 \\
          &       & Turing & 8     & 90.0\% & 0.093 & 160.9 & 8     & 90.0\% & 0.093 & 160.9 & 3     & 3     & 1     & 1     & 1     & 1 \\
          \cline{2-17}
          & Detroit & Archimedes & 11    & 82.5\% & 0.152 & 225.7 & 11    & 82.5\% & 0.152 & 225.7 & 2     & 2     & 1     & 1     & 1     & 1 \\
          &       & Carson & 22    & 85.1\% & 0.124 & 110.2 & 8     & 84.2\% & 0.114 & 105.7 & 2     & 2     & 1     & 1     & 0     & 0 \\
          &       & Curie & 12    & 88.6\% & 0.092 & 151.4 & 4     & 85.1\% & 0.113 & 185.6 & 4     & 4     & 1     & 1     & 1     & 1 \\
          &       & Daly  & 8     & 89.5\% & 0.093 & 118.1 & 8     & 89.5\% & 0.093 & 118.1 & 0     & 0     & 1     & 1     & 0     & 0 \\
          &       & Darwin & 16    & 85.5\% & 0.108 & 96.0  & 11    & 84.6\% & 0.107 & 90.1  & 2     & 2     & 1     & 1     & 1     & 1 \\
          &       & Tesla & 8     & 88.6\% & 0.105 & 134.2 & 8     & 88.6\% & 0.105 & 134.2 & 2     & 2     & 1     & 1     & 1     & 1 \\
\midrule
    Average &       &       & 11.0  & 87.0\% & 0.109 & 143.1 & 8.3   & 86.5\% & 0.110 & 144.4 & 2.6   & 2.6   & 1.1   & 1.0   & 0.7   & 0.7 \\
     \bottomrule
    \end{tabular}%
    }
    \label{tab:HnH-MD}%
\end{table}%

\begin{table}[htbp]
  \tiny
  \centering
  \caption{The numbers of clusters in models selected by the (i) ${\text{MSPE}}_{\text{P}}$ and (ii) ${\text{MSPEB}}_{\text{P}}$  criteria in the TCL method, the $\text{PCP}$, $\text{MSPE}_{\text{P}}$ and $\text{MSPE}_{\text{Y}}$ estimates of these selected optimal models and the numbers of correct predictions of semifinalists, finalists and champions in the playoffs for the 2018 and 2019 FRC championships.}
    \resizebox{\columnwidth}{!}{%
    \begin{tabular}{lllrrrrrrrrrrrrrr}
    \toprule
  &  &  &  (i) & & &  &  (ii) & & & & (i) & (ii) & (i) & (ii) & (i) & (ii)  \\
   Season & Tournament & Division & $\widehat{c}$ & \multicolumn{1}{c}{$\widehat{\textsc{pcp}}$} & \multicolumn{1}{c}{$\widehat{\textsc{mspe}}_{\textsc{p}}$} & \multicolumn{1}{c}{$\widehat{\textsc{mspe}}_{\textsc{y}}$} & $\widetilde{c}$ & \multicolumn{1}{c}{$\widehat{\textsc{pcp}}$} & \multicolumn{1}{c}{$\widehat{\textsc{mspe}}_{\textsc{p}}$} & \multicolumn{1}{c}{$\widehat{\textsc{mspe}}_{\textsc{y}}$} & \multicolumn{2}{l}{Semifinalists}  &\multicolumn{2}{l}{Finalists} & \multicolumn{2}{l}{Champions}  \\
\midrule
    2018  & Houston & Carver & 16    & 88.6\% & 0.084 & 6087.9 & 7     & 86.8\% & 0.092 & 6322.2 & 3     & 3     & 0     & 0     & 1     & 1 \\
          &       & Galileo & 9     & 88.6\% & 0.098 & 7541.4 & 3     & 84.2\% & 0.124 & *12539.2 & 2     & *2     & 2     & *2     & 1     & *1 \\
          &       & Hopper & 9     & 88.6\% & 0.081 & 6557.3 & 8     & 88.6\% & 0.083 & 6447.7 & 3     & 3     & 2     & 2     & 0     & 0 \\
          &       & Newton & 14    & 91.5\% & 0.067 & 5945.8 & 10    & 90.6\% & 0.067 & 5803.8 & 3     & 3     & 2     & 2     & 1     & 0 \\
          &       & Roebling & 8     & 88.8\% & 0.094 & 6027.1 & 6     & 87.9\% & 0.100 & 7393.4 & 2     & 2     & 1     & 1     & 0     & 1 \\
          &       & Turing & 9     & 92.8\% & 0.059 & 6065.3 & 9     & 92.8\% & 0.059 & 6065.3 & 1     & 1     & 1     & 1     & 0     & 0 \\
          \cline{2-17}
          & Detroit & Archimedes & 9     & 87.7\% & 0.100 & 5646.8 & 9     & 87.7\% & 0.100 & 5646.8 & 4     & 4     & 0     & 0     & 1     & 1 \\
          &       & Carson & 13    & 87.3\% & 0.098 & 6125.0 & 8     & 82.9\% & 0.103 & 6116.3 & 4     & 4     & 0     & 1     & 1     & 1 \\
          &       & Curie & 9     & 85.7\% & 0.096 & 4944.3 & 6     & 83.9\% & 0.104 & 6288.5 & 3     & 3     & 2     & 1     & 1     & 1 \\
          &       & Daly  & 8     & 89.9\% & 0.066 & 6522.8 & 7     & 89.9\% & 0.066 & 6967.7 & 3     & 3     & 2     & 1     & 0     & 0 \\
          &       & Darwin & 10    & 85.3\% & 0.112 & 8025.7 & 4     & 80.8\% & 0.134 & *10195.2 & 0     & *0     & 2     & *2     & 0     & *0 \\
          &       & Tesla & 7     & 87.5\% & 0.068 & 5128.3 & 7     & 87.5\% & 0.068 & 5128.3 & 4     & 4     & 0     & 0     & 0     & 0 \\
\midrule
    Average &       &       & 10.1  & 88.5\% & 0.085 & 6218.1 & 7.0   & 87.0\% & 0.092 & 7076.2 & 2.7   & 2.7   & 1.2   & 1.1   & 0.5   & 0.5 \\
\midrule
\midrule
    2019  & Houston & Carver & 11    & 80.7\% & 0.133 & 189.6 & 6     & 77.2\% & 0.139 & 194.8 & 4     & 4     & 1     & 0     & 1     & 1 \\
          &       & Galileo & 9     & 90.6\% & 0.079 & 103.2 & 9     & 90.6\% & 0.079 & 103.2 & 2     & 2     & 2     & 2     & 0     & 0 \\
          &       & Hopper & 15    & 90.4\% & 0.093 & 108.0 & 6     & 84.2\% & 0.097 & 117.4 & 4     & 3     & 1     & 1     & 1     & 1 \\
          &       & Newton & 12    & 83.9\% & 0.107 & 150.0 & 9     & 85.7\% & 0.112 & 178.7 & 3     & 3     & 1     & 1     & 1     & 1 \\
          &       & Roebling & 5     & 84.4\% & 0.124 & 130.2 & 4     & 84.4\% & 0.124 & 129.0 & 3     & 3     & 2     & 2     & 0     & 0 \\
          &       & Turing & 7     & 88.2\% & 0.090 & 154.2 & 7     & 88.2\% & 0.090 & 154.2 & 1     & 1     & 2     & 2     & 1     & 1 \\
          \cline{2-17}
          & Detroit & Archimedes & 12    & 80.7\% & 0.128 & 159.1 & 5     & 72.8\% & 0.157 & 266.1 & 2     & 2     & 1     & 1     & 1     & 1 \\
          &       & Carson & 10    & 83.3\% & 0.108 & 96.2  & 7     & 83.3\% & 0.118 & 113.8 & 2     & 2     & 1     & 1     & 0     & 0 \\
          &       & Curie & 10    & 86.8\% & 0.087 & 144.1 & 10    & 86.8\% & 0.087 & 144.1 & 3     & 3     & 1     & 1     & 1     & 1 \\
          &       & Daly  & 8     & 89.5\% & 0.093 & 118.1 & 8     & 89.5\% & 0.093 & 118.1 & 0     & 0     & 1     & 1     & 0     & 0 \\
          &       & Darwin & 11    & 84.6\% & 0.107 & 90.1  & 11    & 84.6\% & 0.107 & 90.1  & 2     & 2     & 1     & 1     & 1     & 1 \\
          &       & Tesla & 9     & 87.7\% & 0.099 & 130.7 & 6     & 86.0\% & 0.112 & 161.1 & 2     & 2     & 1     & 1     & 1     & 1 \\
\midrule
    Average &       &       & 9.9   & 85.9\% & 0.104 & 131.1 & 7.3   & 84.4\% & 0.110 & 147.6 & 2.3   & 2.3   & 1.3   & 1.2   & 0.7   & 0.7 \\
     \bottomrule
    \end{tabular}%
    }
    \label{tab:HnH-MP}%
\end{table}%

\begin{table}[htbp]
  \tiny
  \centering
  \caption{The numbers of clusters in models selected by the (i) ${\text{MSPE}}_{\text{Y}}$ and (ii) ${\text{MSPEB}}_{\text{Y}}$  criteria in the TCL method, the $\text{PCP}$, $\text{MSPE}_{\text{P}}$ and $\text{MSPE}_{\text{Y}}$ estimates of these selected optimal models and the numbers of correct predictions of semifinalists, finalists and champions in the playoffs for the 2018 and 2019 FRC championships.}
    \resizebox{\columnwidth}{!}{%
    \begin{tabular}{lllrrrrrrrrrrrrrr}
    \toprule
  &  &  &  (i) & & &  &  (ii) & & & & (i) & (ii) & (i) & (ii) & (i) & (ii)  \\
    Season & Tournament & Division & $\widehat{c}$ & \multicolumn{1}{c}{$\widehat{\textsc{pcp}}$} & \multicolumn{1}{c}{$\widehat{\textsc{mspe}}_{\textsc{p}}$} & \multicolumn{1}{c}{$\widehat{\textsc{mspe}}_{\textsc{y}}$} & $\widetilde{c}$ & \multicolumn{1}{c}{$\widehat{\textsc{pcp}}$} & \multicolumn{1}{c}{$\widehat{\textsc{mspe}}_{\textsc{p}}$} & \multicolumn{1}{c}{$\widehat{\textsc{mspe}}_{\textsc{y}}$} & \multicolumn{2}{l}{Semifinalists}  &\multicolumn{2}{l}{Finalists} & \multicolumn{2}{l}{Champions}  \\
\midrule
    2018  & Houston & Carver & 10    & 86.0\% & 0.092 & 5584.7 & 7     & 86.8\% & 0.092 & 6322.2 & 3     & 3     & 0     & 0     & 1     & 1 \\
          &       & Galileo & 12    & 86.0\% & 0.102 & 7312.1 & 9     & 88.6\% & 0.098 & 7541.4 & 3     & 2     & 2     & 2     & 1     & 1 \\
          &       & Hopper & 8     & 88.6\% & 0.083 & 6447.7 & 8     & 88.6\% & 0.083 & 6447.7 & 3     & 3     & 2     & 2     & 0     & 0 \\
          &       & Newton & 8     & 89.7\% & 0.076 & 5715.7 & 8     & 89.7\% & 0.076 & 5715.7 & 3     & 3     & 2     & 2     & 0     & 0 \\
          &       & Roebling & 8     & 88.8\% & 0.094 & 6027.1 & 8     & 88.8\% & 0.094 & 6027.1 & 2     & 2     & 1     & 1     & 0     & 0 \\
          &       & Turing & 9     & 92.8\% & 0.059 & 6065.3 & 9     & 92.8\% & 0.059 & 6065.3 & 1     & 1     & 1     & 1     & 0     & 0 \\
          \cline{2-17}
          & Detroit & Archimedes & 9     & 87.7\% & 0.100 & 5646.8 & 9     & 87.7\% & 0.100 & 5646.8 & 4     & 4     & 0     & 0     & 1     & 1 \\
          &       & Carson & 11    & 83.8\% & 0.102 & 6079.9 & 8     & 82.9\% & 0.104 & 6116.3 & 4     & 4     & 1     & 1     & 1     & 1 \\
          &       & Curie & 9     & 85.7\% & 0.096 & 4944.3 & 9     & 85.7\% & 0.096 & 4944.3 & 3     & 3     & 2     & 2     & 1     & 1 \\
          &       & Daly  & 11    & 91.7\% & 0.067 & 6408.7 & 8     & 89.9\% & 0.066 & 6522.8 & 3     & 3     & 2     & 2     & 0     & 0 \\
          &       & Darwin & 11    & 86.2\% & 0.115 & 7864.8 & 5     & 79.9\% & 0.132 & 8723.2 & 0     & 0     & 1     & 2     & 0     & 1 \\
          &       & Tesla & 7     & 87.5\% & 0.068 & 5128.3 & 7     & 87.5\% & 0.068 & 5128.3 & 4     & 4     & 0     & 0     & 0     & 0 \\
\midrule
    Average &       &       & 9.4   & 87.9\% & 0.088 & 6102.1 & 7.9   & 87.4\% & 0.089 & 6266.8 & 2.8   & 2.7   & 1.2   & 1.3   & 0.4   & 0.5 \\
\midrule
\midrule
    2019  & Houston & Carver & 10    & 81.6\% & 0.136 & 186.2 & 6     & 77.2\% & 0.139 & 194.8 & 4     & 4     & 1     & 0     & 1     & 1 \\
          &       & Galileo & 9     & 90.6\% & 0.078 & 103.2 & 9     & 90.6\% & 0.079 & 103.2 & 2     & 2     & 2     & 2     & 0     & 0 \\
          &       & Hopper & 10    & 89.5\% & 0.097 & 99.3  & 5     & 86.0\% & 0.108 & 118.3 & 4     & 4     & 0     & 1     & 1     & 0 \\
          &       & Newton & 11    & 83.9\% & 0.107 & 147.5 & 10    & 84.8\% & 0.109 & 151.7 & 3     & 3     & 1     & 1     & 1     & 1 \\
          &       & Roebling & 10    & 76.3\% & 0.135 & 115.1 & 4     & 84.4\% & 0.124 & 129.0 & 3     & 3     & 2     & 2     & 0     & 0 \\
          &       & Turing & 9     & 88.2\% & 0.091 & 143.0 & 7     & 88.2\% & 0.090 & 154.2 & 1     & 1     & 2     & 2     & 1     & 1 \\
          \cline{2-17}
          & Detroit & Archimedes & 12    & 80.7\% & 0.128 & 159.1 & 12    & 80.7\% & 0.128 & 159.1 & 2     & 2     & 1     & 1     & 1     & 1 \\
          &       & Carson & 11    & 82.5\% & 0.113 & 93.0  & 9     & 83.3\% & 0.118 & 99.8  & 2     & 2     & 1     & 1     & 0     & 0 \\
          &       & Curie & 10    & 86.8\% & 0.087 & 144.1 & 8     & 85.1\% & 0.104 & 151.2 & 3     & 4     & 1     & 0     & 1     & 1 \\
          &       & Daly  & 9     & 86.0\% & 0.095 & 115.9 & 6     & 84.2\% & 0.105 & 119.6 & 0     & 1     & 1     & 1     & 0     & 0 \\
          &       & Darwin & 11    & 84.6\% & 0.107 & 90.1  & 11    & 84.6\% & 0.107 & 90.1  & 2     & 2     & 1     & 1     & 1     & 1 \\
          &       & Tesla & 9     & 87.7\% & 0.099 & 130.7 & 7     & 87.7\% & 0.109 & 132.7 & 2     & 2     & 1     & 1     & 1     & 1 \\
\midrule
    Average &       &       & 10.1  & 84.9\% & 0.106 & 127.3 & 7.8   & 84.7\% & 0.110 & 133.6 & 2.3   & 2.5   & 1.2   & 1.1   & 0.7   & 0.6 \\
     \bottomrule
    \end{tabular}%
    }
  \label{tab:HnH-MY}%
\end{table}%

\begin{table}[htbp]
  \tiny
  \centering
  \caption{The number of robots, the $\text{PCP}$, $\text{MSPE}_{\text{P}}$ and $\text{MSPE}_{\text{Y}}$ estimates of the WMPR model and the numbers of correct predictions of semifinalists, finalists and champions in the playoffs for the 2018 and 2019 FRC championships.}
      \resizebox{!}{3.73cm}{%
    \begin{tabular}{lllrrrrrrrr}
    \toprule    
     Season & Tournament & Division & $K$ & \multicolumn{1}{c}{$\widehat{\textsc{pcp}}$} & \multicolumn{1}{c}{$\widehat{\textsc{mspe}}_{\textsc{p}}$} & \multicolumn{1}{c}{$\widehat{\textsc{mspe}}_{\textsc{y}}$} & \multicolumn{1}{c}{Semifinalists}  &\multicolumn{1}{c}{Finalists} & \multicolumn{1}{c}{Champions}  \\
\midrule
    2018  & Houston & \multicolumn{1}{l}{Carver} & 68    & 73.7\% & 0.229 & 26782.8 & 3     & 0     & 1 \\
          &       & \multicolumn{1}{l}{Galileo} & 68    & 70.2\% & 0.241 & 33306.8 & 2     & 2     & 1 \\
          &       & \multicolumn{1}{l}{Hopper} & 68    & 69.3\% & 0.233 & 30922.8 & 3     & 2     & 0 \\
          &       & \multicolumn{1}{l}{Newton} & 67    & 76.3\% & 0.178 & 25571.7 & 3     & 2     & 0 \\
          &       & \multicolumn{1}{l}{Roebling} & 67    & 68.3\% & 0.251 & 28662.4 & 2     & 1     & 0 \\
          &       & \multicolumn{1}{l}{Turing} & 67    & 73.9\% & 0.199 & 30199.0 & 2     & 1     & 0 \\
          \cline{2-10}
          & Detroit & \multicolumn{1}{l}{Archimedes} & 68    & 63.2\% & 0.248 & 26676.4 & 4     & 0     & 1 \\
          &       & \multicolumn{1}{l}{Carson} & 68    & 64.5\% & 0.275 & 28789.7 & 4     & 0     & 1 \\
          &       & \multicolumn{1}{l}{Curie} & 67    & 65.2\% & 0.277 & 23039.6 & 3.5   & 2     & 1 \\
          &       & \multicolumn{1}{l}{Daly} & 68    & 69.7\% & 0.216 & 29722.2 & 2     & 1     & 0 \\
          &       & \multicolumn{1}{l}{Darwin} & 67    & 63.8\% & 0.295 & 36873.5 & 0     & 1     & 0 \\
          &       & \multicolumn{1}{l}{Tesla} & 67    & 76.8\% & 0.189 & 25379.2 & 3     & 0     & 0 \\
             \midrule
    Average &       &       & 67.5  & 69.6\% & 0.236 & 28827.2 & 2.6   & 1.0   & 0.4 \\
        \midrule
        \midrule
    2019  & Houston & Carver & 68    & 62.3\% & 0.312 & 912.2 & 4     & 1     & 1 \\
          &       & Galileo & 67    & 70.1\% & 0.224 & 496.2 & 2     & 2     & 0 \\
          &       & Hopper & 68    & 67.5\% & 0.257 & 469.6 & 4     & 1     & 1 \\
          &       & Newton & 67    & 69.6\% & 0.257 & 735.3 & 3     & 1     & 1 \\
          &       & Roebling & 67    & 65.6\% & 0.283 & 559.0 & 3     & 2     & 0 \\
          &       & Turing & 66    & 70.9\% & 0.248 & 679.3 & 1     & 2     & 1 \\
          \cline{2-10}
          & Detroit & Archimedes & 68    & 64.9\% & 0.292 & 738.3 & 2     & 1     & 1 \\
          &       & Carson & 68    & 68.4\% & 0.253 & 425.9 & 2     & 1     & 0 \\
          &       & Curie & 68    & 71.1\% & 0.228 & 622.9 & 4     & 0     & 1 \\
          &       & Daly  & 68    & 63.2\% & 0.266 & 556.4 & 0     & 1     & 0 \\
          &       & Darwin & 68    & 69.7\% & 0.222 & 388.2 & 1     & 1     & 1 \\
          &       & Tesla & 68    & 68.4\% & 0.233 & 609.8 & 2     & 1     & 1 \\
    \midrule
    Average &       &       & 67.6  & 67.6\% & 0.256 & 599.4 & 2.3   & 1.2   & 0.7 \\
     \bottomrule
    \end{tabular}%
    }
  \label{tab:WMPR}%
\end{table}%

As shown in Tables \ref{tab:WMPRC2-MY}, \ref{tab:HnH-MY} and \ref{tab:WMPR}, models selected by the $\text{MSPE}_{\text{Y}}$ criterion in both the LCT and TCL methods generally have better predictions of semifinalists, finalists and champions in the playoffs than the WMPR model.
Compared to models selected by any one of the MSPE criteria in the LCT method, models selected by the same criterion in the TCL method have comparable or even better predictive performance for the 2018 and 2019 FRC championships.
For models selected by one of the $\text{MSPE}_{\text{D}}$ and $\text{MSPE}_{\text{Y}}$ criteria in the TCL method, those selected by the corresponding MSPEB criteria have better predictive capacities for the 2018 FRC championships but relatively poor predictive capacities for the 2019 FRC championships.
Since models selected by the $\text{MSPEB}_{\text{P}}$ criterion in the TCL method are accompanied by extremely large $\text{MSPE}_{\text{Y}}$ estimates for the Galileo and Darwin divisions of the 2018 FRC championships, the $\text{MSPE}_{\text{P}}$ criterion was used to select optimal models in these two divisions.
For the 2019 FRC championships, models selected by the $\text{MSPE}_{\text{P}}$ criterion also outperform those selected by the $\text{MSPEB}_{\text{P}}$ criterion.
Overall, the findings are consistent with the simulation results in the next subsection.

By borrowing a rule of \cite{HLS2013}, clusters of two selected optimal models are regarded to have an outstanding nested relation for $\text{MINR} \geq 0.9$; an excellent nested relation for $0.8 \leq \text{MINR} < 0.9$; an acceptable nested relation for $0.7 \leq \text{MINR} < 0.8$; and a poor nested relation for $\text{MINR}<0.7$.
It is further used to classify the strength of the RC for the monotonic association between robot strengths of two selected optimal models.
Figures \ref{fig:MINR} and \ref{fig:RC} display, respectively, the averages of MINR's and RC's over the divisions in Houston and Detroit.
There exist outstanding MINR's of models selected by the MSPE criteria in the LCT method for the 2018 and 2019 FRC championships.
These values are further found to be higher than those of models, which have outstanding or excellent MINR's, selected by the criteria in the TCL method.
Moreover, the MINR of models selected by any two of the MSPE criteria in the LCT method for the 2018 FRC championships is generally higher than that of models selected by the same criteria for the 2019 FRC championships.
This conclusion can also be made for the MINR of models selected by the same MSPE criterion in both the LCT and TCL methods for the 2018 and 2019 FRC championships.
For the 2019 FRC championships, clusters of some selected optimal models have only acceptable nested relations.
Figures \ref{fig:MINR} and \ref{fig:RC} show that robot strengths from any two models with an outstanding or excellent MINR have outstanding or excellent monotonic associations.
Any two selected optimal models with an acceptable MINR are further found to have an excellent RC.
As for models selected by any two criteria in either the LCT method or the TCL method, the RC for the 2018 FRC championships is generally higher than that for the 2019 FRC championships.
\begin{figure}[htbp]
 \centering
 \includegraphics[scale=0.43]{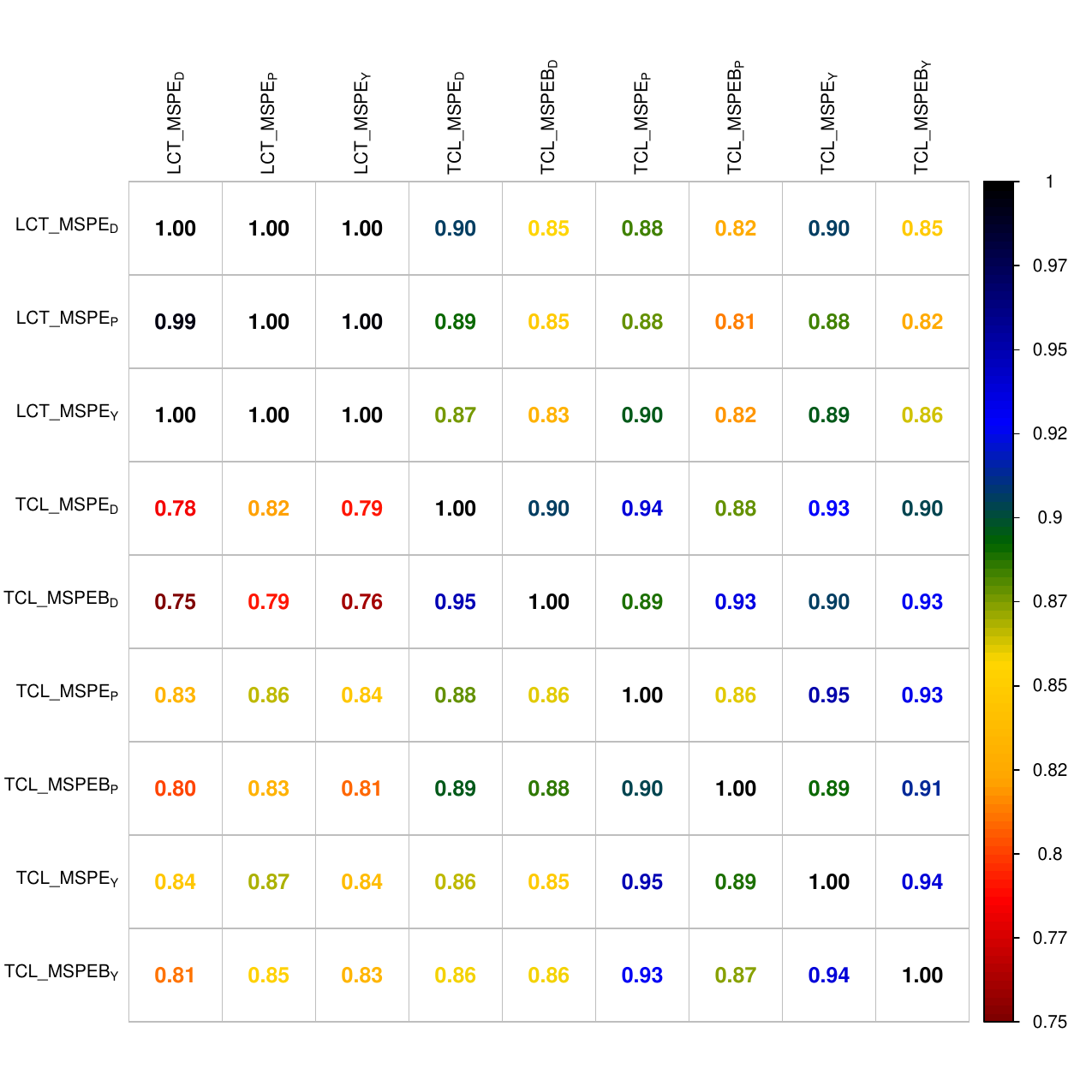}
 \caption{{\footnotesize The averages of the MINR's of two selected optimal models over the divisions in the Houston and Detroit tournaments of the 2018 (upper triangular part) and 2019 (lower triangular part) FRC championships.} }
 \label{fig:MINR}
\end{figure}

\begin{figure}[htbp]
 \centering
 \includegraphics[scale=0.43]{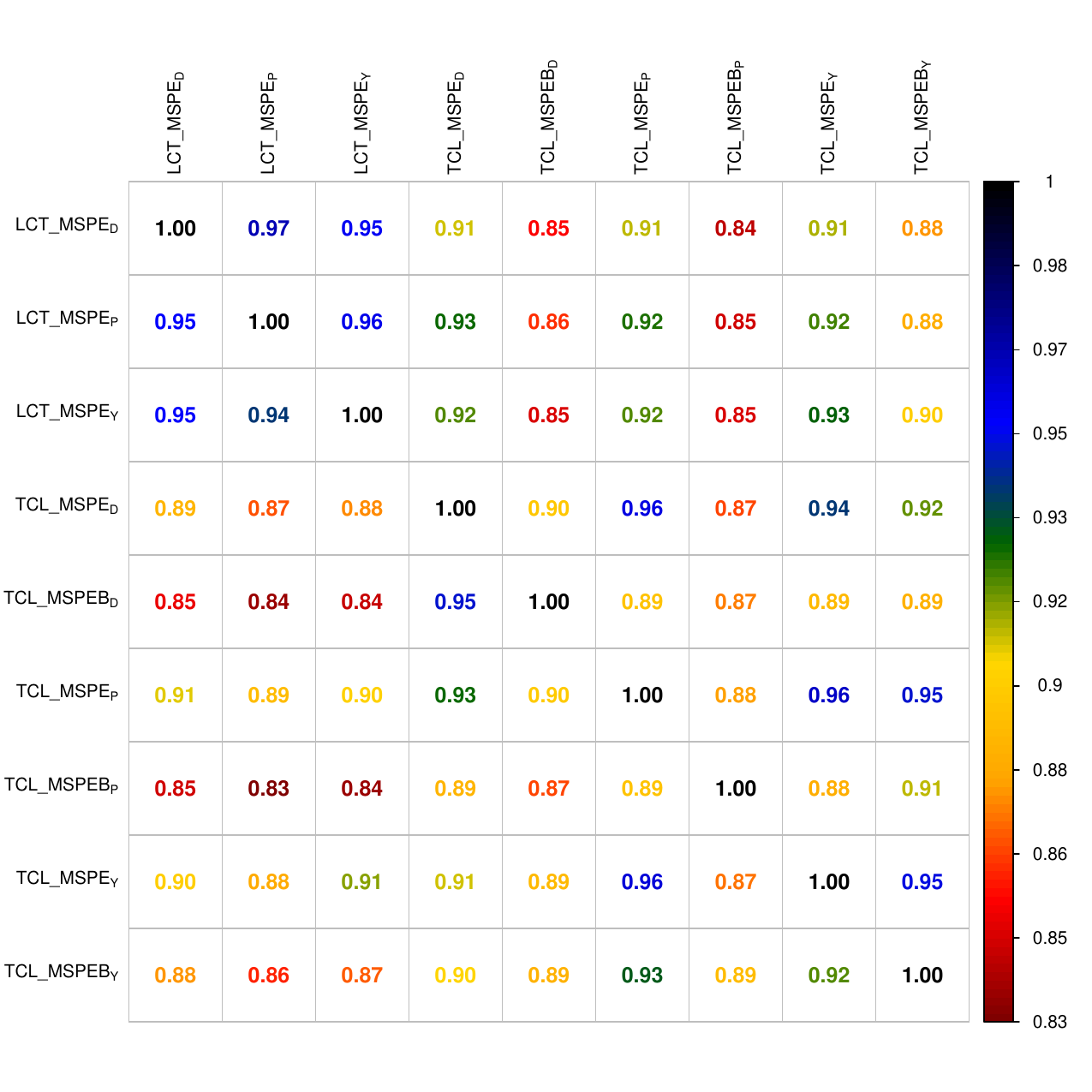}
 \caption{{\footnotesize The averages of the RC's of two selected optimal models over the divisions in the Houston and Detroit tournaments of the 2018 (upper triangular part) and 2019 (lower triangular part) FRC championships.} }
 \label{fig:RC}
\end{figure}
\end{subsection}

\begin{subsection}{Monte Carlo Simulation}
In this simulation study, data were similar in nature to qualification stage data for the divisions of the 2019 Houston and Detroit championships.
We designed $c_o$, $\beta^o$, $\sigma^2$ and $X$ in the proposed WMPRC model based on $\widehat{c}$, $\widehat{\beta}^{\widehat{c}}$, $\widehat{\sigma}^2 = {\widehat{\textsc{mspe}}_{\textsc{y}}\big(\widehat{c},g^{\widehat{c}}\big)}$ and $X$ from the following selected optimal models:
\begin{description}
\item{M1.} The model, which is the same as those selected by the $\text{MSPEB}_{\text{P}}$ and $\text{MSPEB}_{\text{Y}}$ criteria, is selected by the $\text{MSPEB}_{\text{D}}$ criterion in the TCL method for the Roebling division with $\widehat{c} = 4$, $\big(\widehat{\beta}^4_{1}, \dots, \widehat{\beta}^4_{4}\big) = (-15.07, -4.75, 4.76, 14.52)$, $\big(\big|G^4_1\big|, \dots, \big|G^4_4\big|\big) = (9, 25, 24, 9)$ and $\widehat{\sigma} = 11.056$.
\item{M2.} The model, which is the same as that selected by the $\text{MSPEB}_{\text{P}}$ criterion, is selected by the $\text{MSPEB}_{\text{D}}$ criterion in the TCL method for the Daly division with $\widehat{c} = 8$, $\big(\widehat{\beta}^8_{1}, \dots, \widehat{\beta}^8_{8}\big) = (-18.16, -13.57$, $-9.81, -0.58, 4.88, 7.90, 11.16, 18.20)$, $\big(\big|G^8_1\big|, \dots, \big|G^8_8\big|\big) = (5, 3, 8, 22$, $17, 6, 5, 2)$ and $\widehat{\sigma} = 10.275$.
\end{description}
For models M1 and M2, the corresponding ratios of the differences between adjacent estimated robot strengths to $\widehat{\sigma}$ are $(0.93, 0.86,$ $0.88)$ and $(0.45, 0.37, 0.9, 0.53, 0.29$, $0.31, 0.68)$.
The error was further specified to follow a normal distribution with mean zero and variance $\sigma^2$.
With this setup, the difference in scores $Y$ was generated from model (\ref{eq:lnmodel}) with the designed model M1 or M2 for $X \beta^o$ and one of the values in $\{0.25 \widehat{\sigma}, 0.5 \widehat{\sigma}, \widehat{\sigma}, 2 \widehat{\sigma}, 4 \widehat{\sigma} \}$ for $\sigma$.
To ensure the numerical stability, the results were based on 5000 replications.

In both the LCT and TCL methods, the MSPE measures of model M1 are almost the same as the minima of the corresponding MSPE curves for $ \sigma \leq 5.528$ (Figure \ref{plot:Roebling_mspe}), whereas the MSPE measures of model M2 are close to the minima of the corresponding MSPE curves only for $\sigma = 2.569$ (Figure \ref{plot:Daly_mspe}).
This is mainly because the proportion of sufficiently separated robot strengths in model M1 is higher than that in model M2.
As expected, the deviations between the MSPE measures of models M1 and M2 and the minima of the corresponding MSPE curves become larger as $\sigma$ increases.
Except for fairly large numbers of clusters, it can be seen in Figures \ref{plot:Roebling_mspe} -- \ref{plot:Daly_mspeb} that the MSPE and MSPEB curves are higher than their estimated curves under models M1 and M2.
With any one of the MSPEB criteria in the TCL method, the MSPE measures of selected optimal models approach the minima of the corresponding MSPE curves.
The minimizers of the MSPEB curves are also smaller than those of their estimated curves and the corresponding MSPE curves.
The parameters in their selected optimal models may be very different, albeit each estimated MSPE curve in the LCT method is close to that in the TCL method.
Especially, the MSPE measures and estimates of models M1 and M2 are, respectively, smaller and larger than the corresponding MSPE measures and estimates of selected optimal models.
Therefore, these estimates are inappropriate to assess the goodness-of-fit for selected optimal models.

\begin{figure}[htb]
\centering
 \includegraphics[width=14cm,height=9.5cm]{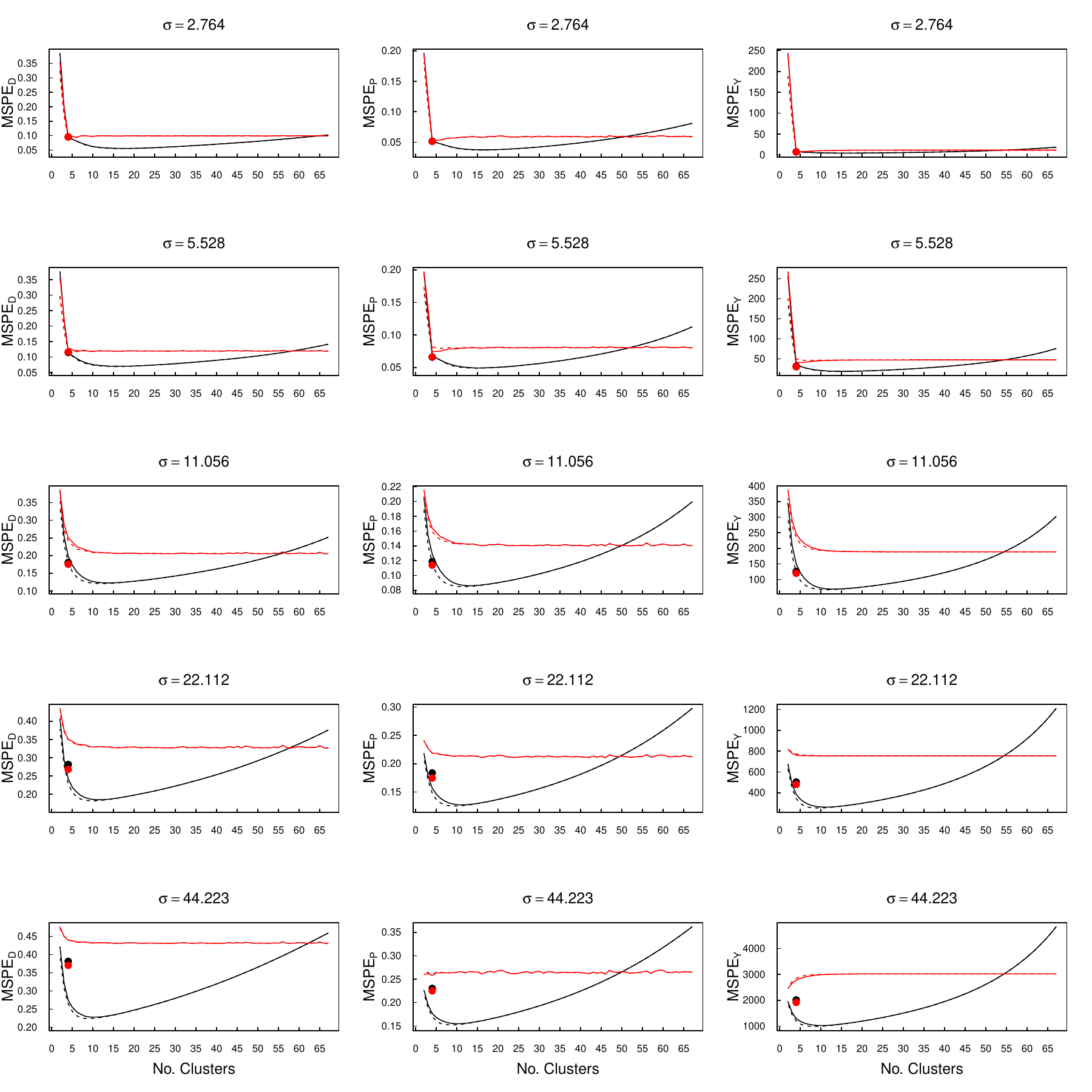}
 \caption{{\footnotesize The MSPE curves and the averages of their 5000 estimated curves in both the LCT and TCL methods (LCT: red and black dashed lines; TCL: red and black solid lines) under model M1. The red and black dots, respectively, denote the MSPE measures of model M1 and the averages of their 5000 estimates.}} \label{plot:Roebling_mspe} 
\end{figure}

\begin{figure}[htbp]
\centering
 \includegraphics[width=14cm,height=9.5cm]{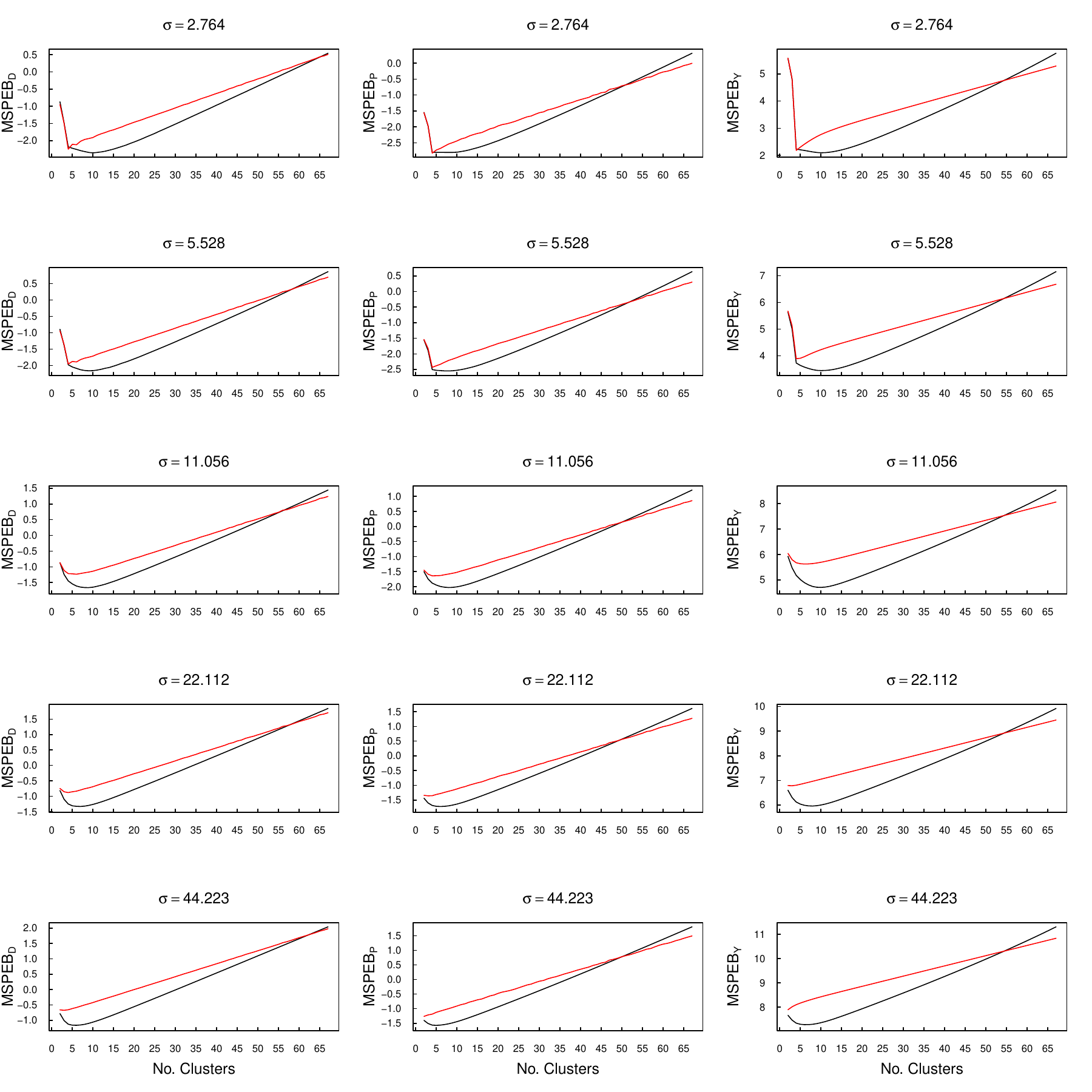}
 \caption{{\footnotesize The MSPEB curves (red line) and the averages of their 5000 estimated curves (black line) in the TCL method under model M1.}} \label{plot:Roebling_mspeb} 
\end{figure}

\begin{figure}[htbp]
\centering
 \includegraphics[width=14cm,height=9.5cm]{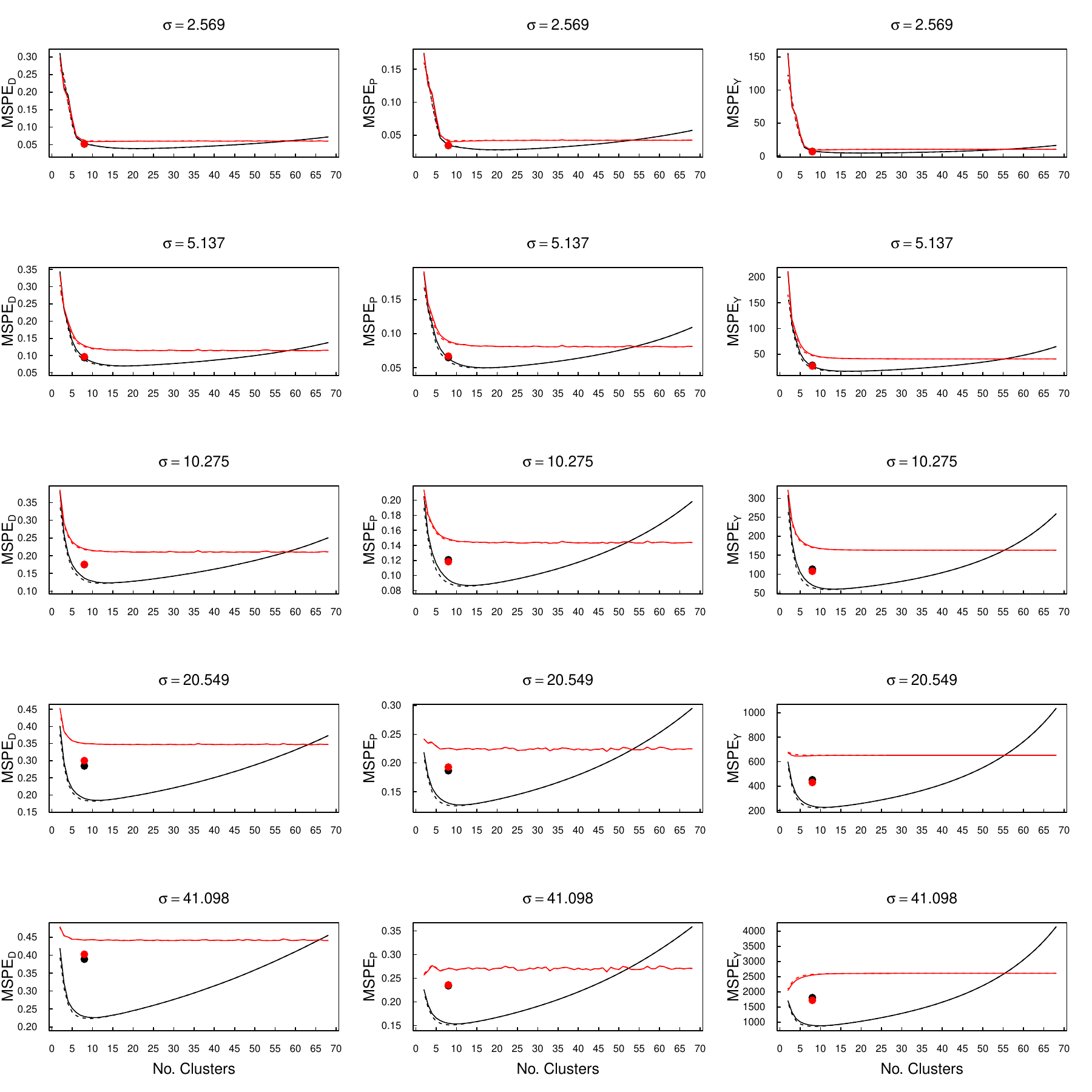}
 \caption{{\footnotesize The MSPE curves and the averages of their 5000 estimated curves in both the LCT and TCL methods (LCT: red and black dashed lines; TCL: red and black solid lines) under model M2. The red and black dots, respectively, denote the MSPE measures of model M2 and the averages of their 5000 estimates.}} \label{plot:Daly_mspe} 
\end{figure}

\begin{figure}[htbp]
\centering
 \includegraphics[width=14cm,height=9.5cm]{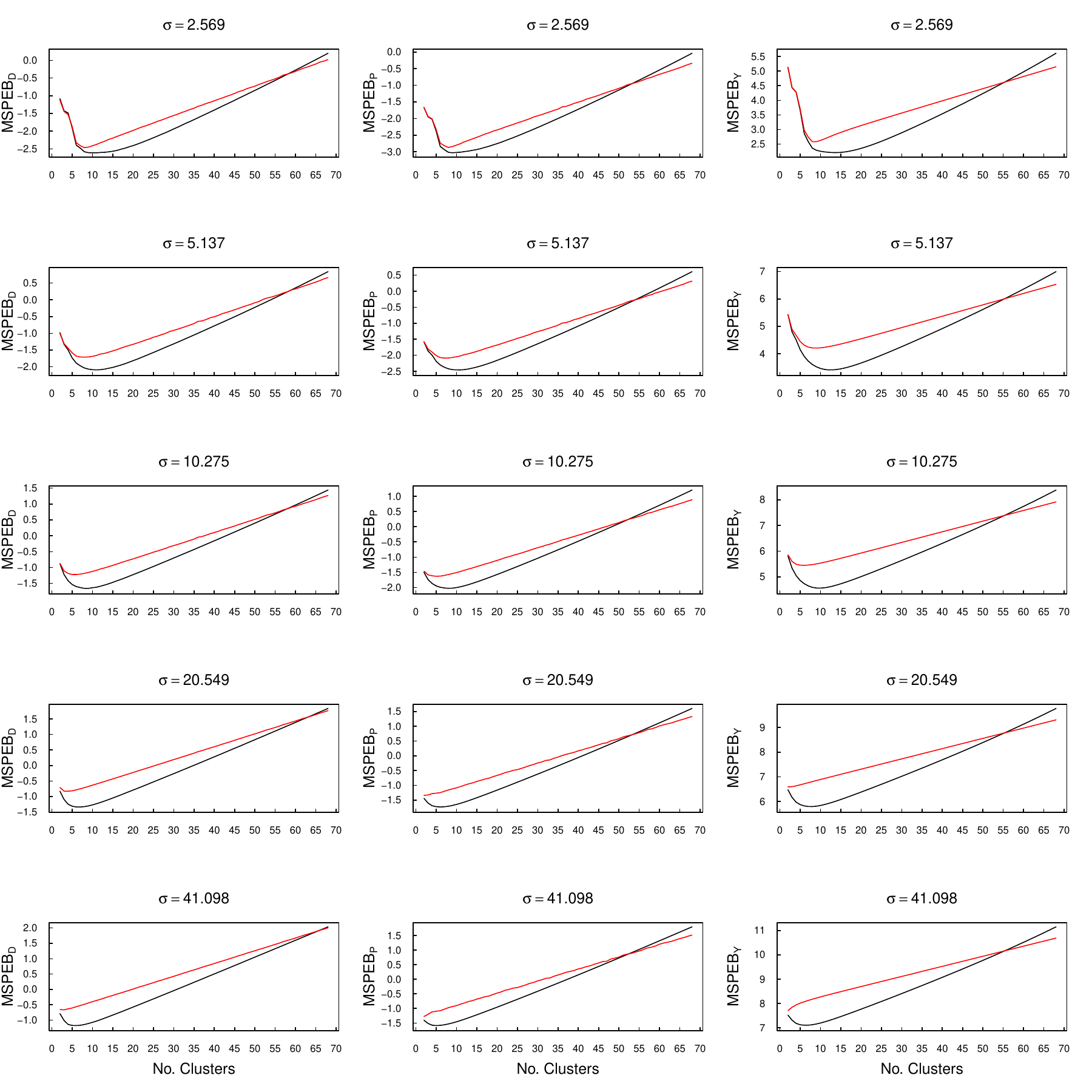}
 \caption{{\footnotesize  The MSPEB curves (red line) and the averages of their 5000 estimated curves (black line) in the TCL method under model M2.}} \label{plot:Daly_mspeb} 
\end{figure}

With the same MSPE criterion in both the LCT and TCL methods, their estimated numbers of clusters in Table \ref{tab:simu_c} are quite close to each other.
The number of clusters in a model selected by the $\text{MSPE}_{\text{D}}$ criterion is further found to be smaller than those  in models selected by the $\text{MSPE}_{\text{P}}$ and $\text{MSPE}_{\text{Y}}$ criteria, which have similar performance in estimating the number of clusters.
In the TCL method, the numbers of clusters in models selected by the MSPEB criteria are shown to be noticeably smaller than those in models selected by the corresponding MSPE criteria.
Moreover, the $\text{MSPEB}_{\text{P}}$ criterion produces the smallest number of clusters and both the $\text{MSPEB}_{\text{D}}$ and $\text{MSPEB}_{\text{Y}}$ criteria have comparable numbers of clusters.
As we can see, the number of clusters in model M1 is overestimated by all the model selection criteria but that in model M2 is only overestimated by the MSPE criteria.
Due to a relatively low proportion of sufficiently separated robot strengths in model M2, the number of clusters is underestimated by the $\text{MSPEB}_{\text{D}}$ and $\text{MSPEB}_{\text{Y}}$ criteria for $\sigma \geq 20.549$ and by the $\text{MSPEB}_{\text{P}}$ criterion for $\sigma \geq 10.275$.
To achieve the minimal $\text{MSPE}_{\text{D}}$, $\text{MSPE}_{\text{P}}$, or $\text{MSPE}_{\text{Y}}$, this underestimation is attributed to the increase of biases and decrease of variances of estimated robot strengths.

\begin{table}[htbp]
\tiny
  \centering
  \caption{The means (standard deviations) of 5000 estimated numbers of clusters in both the LCT and TCL methods under models M1 and M2.}
  \resizebox{0.9\columnwidth}{3.2cm}{%
    \begin{tabular}{crrrrrrrrrrr}
    \toprule
    Model & \multicolumn{1}{c}{$\sigma$}   & \multicolumn{3}{c}{LCT method} & & \multicolumn{6}{c}{TCL method}\\
    \cline{3-5} \cline{7-12}
         &    & \multicolumn{1}{c}{$\textsc{mspe}_{\textsc{d}}$}  & \multicolumn{1}{c}{$\textsc{mspe}_{\textsc{p}}$} &  \multicolumn{1}{c}{$\textsc{mspe}_{\textsc{y}}$} &       & \multicolumn{1}{c}{$\textsc{mspe}_{\textsc{d}}$} & \multicolumn{1}{c}{$\textsc{mspeb}_{\textsc{d}}$} & \multicolumn{1}{c}{$\textsc{mspe}_{\textsc{p}}$} & \multicolumn{1}{c}{$\textsc{mspeb}_{\textsc{p}}$} & \multicolumn{1}{c}{$\textsc{mspe}_{\textsc{y}}$} & \multicolumn{1}{c}{$\textsc{mspeb}_{\textsc{y}}$} \\
         \midrule
    M1 & 2.764    & 13.0    & 14.9    & 15.1   &       & 13.1  & 10.4  & 14.7  & 8.7   & 15.1  & 9.6 \\
          &     & (5.46)    & (5.38)   & (2.23)   &       & (5.10)  & (4.06)  & (5.08)  & (3.45)  & (2.60)  & (2.15) \\
    & 5.528  & 12.4     & 14.0    & 14.3   &       & 12.5  & 9.4   & 13.7  & 7.9   & 14.4  & 9.4 \\
          &     & (5.09)   & (4.79)   & (1.96)   &       & (4.85)  & (3.58)  & (4.51)  & (3.00)  & (2.39)  & (2.06) \\
     & 11.056  & 10.5    & 11.2    & 11.2   &       & 11.2  & 8.0   & 11.6  & 7.4   & 11.7  & 8.6 \\
          &      & (4.11)   & (3.35)   & (1.61)    &       & (3.92)  & (2.67)  & (3.26)  & (2.24)  & (2.16)  & (1.98) \\
    & 22.112  & 8.9      & 9.1      & 9.0      &       & 9.6   & 6.5   & 9.7   & 5.7   & 9.8   & 6.9 \\
          &      & (3.47)    & (2.57)    & (1.47)   &       & (3.44)  & (2.26)  & (2.75)  & (1.87)  & (2.10)  & (1.81) \\
    & 44.223   & 8.2   & 8.2     & 8.1 &       & 8.7   & 5.7   & 8.8   & 4.9   & 9.0   & 6.0 \\
          &      & (3.08) & (2.32) & (1.39)   &       & (3.18)  & (2.07)  & (2.54)  & (1.64)  & (2.03)  & (1.74) \\
          \midrule \midrule
    M2 & 2.569   & 15.1    & 18.3   & 19.2    &       & 15.2  & 12.4  & 17.7  & 11.2  & 19.2  & 12.7 \\
          &     & (6.08)  & (6.42)    & (2.43)    &       & (5.78)  & (4.69)  & (5.96)  & (3.87)  & (2.71)  & (2.54) \\
    & 5.137  & 13.8    & 15.4   & 15.6    &       & 14.1  & 10.9  & 15.3  & 9.9   & 15.9  & 11.3 \\
          &     & (5.10)   & (4.72)    & (2.00)    &       & (4.90)  & (3.60)  & (4.51)  & (2.81)  & (2.46)  & (2.19) \\
     & 10.275 & 11.1   & 11.7    & 11.7    &       & 11.6  & 8.2   & 11.9  & 7.4   & 12.1  & 8.7 \\
          &     & (4.16)    & (3.42)    & (1.69)    &       & (4.16)  & (2.73)  & (3.39)  & (2.21)  & (2.16)  & (1.90) \\
    & 20.549&  9.2     & 9.3    & 9.3     &       & 9.6   & 6.6   & 9.8   & 5.8   & 10.0  & 6.9 \\
          &     & (3.50)   & (2.60)    & (1.51)    &       & (3.44)  & (2.23)  & (2.73)  & (1.80)  & (2.04)  & (1.71) \\
    & 41.098  & 8.2     & 8.3   & 8.2    &       & 8.7   & 5.8   & 8.8   & 4.9   & 8.8   & 5.9 \\
          &     & (3.19)    & (2.35) & (1.40)    &       & (3.21)  & (2.01)  & (2.46)  & (1.57)  & (1.94)  & (1.58) \\
               \bottomrule
    \end{tabular}%
    }
  \label{tab:simu_c}%
\end{table}%

Tables \ref{tab:simu_mse} -- \ref{tab:simu_mspey} summarize the performance of the WMPR model in the estimation of robot strengths and prediction of match scores (or outcomes).
In Table \ref{tab:simu_mse}, the mean squared error (MSE) of estimated robot strengths from the WMPR model is significantly improved by the TCL method with the $\text{MSPEB}_{\text{P}}$ criterion for $\sigma \leq 5.528$ or $\sigma \geq 22.112$ under model M1 and for $\sigma = 2.569$ or $\sigma \geq 20.549$ under model M2.
The MSE of estimated robot strengths from the WMPR model is, however, slightly smaller than that of estimated robot strengths 
from a model selected by the $\text{MSPE}_{\text{Y}}$ criterion in the TCL method for $\sigma = 11.056$ under model M1 and for $5.137 \leq \sigma \leq 10.275$ under model M2.
In an attempt to alleviate overparameterization of the WMPR model, the gain in identifying clusters of robot strengths still exceeds the loss in introducing additional biases and variances of estimated number of clusters and clusters of robots.
The same conclusion can be further made for the MINR's in Table \ref{tab:simu_minr}.
Compared to the RC of the true model and a model selected by the best criterion in the TCL method, the RC of the true model and the WMPR model (see Table \ref{tab:simu_rc}) is notably lower for $\sigma \leq 11.056$ under model M1 and for $\sigma \leq 5.137$ under model M2; slightly lower for $\sigma = 22.112$ under model M1 and for $\sigma = 10.275$ under model M2; slightly higher for $\sigma = 44.223$ under model M1 and for $\sigma \geq 20.549$ under model M2.
To reduce the MSE of estimated robot strengths from the WMPR model for the case of extremely large values of $\sigma$, the established criteria in the TCL method tend to increase the bias and decrease the variance of estimated robot strengths from a selected optimal model.
As a result, it will lower the RC of the true model and a selected optimal model.
In Tables \ref{tab:simu_msped} -- \ref{tab:simu_mspey}, the MSPE measures of the WMPR model are found to be overestimated for all values of $\sigma$ under models M1 and M2.
The conclusion for the MSE of estimated robot strengths can also be made for the predictive performance of the WMPR model and our selected optimal models in terms of the MSPE measures.

With the same MSPE criterion for model selection, Table \ref{tab:simu_mse} shows that estimated robot strengths in the LCT method can be improved by the proposed classification system in terms of the MSE.
In the TCL method, the MSE of estimated robot strengths from a model selected by any one of the MSPEB criteria is smaller than that of estimated robot strengths from a model selected by the corresponding MSPE criterion for $\sigma \leq 5.528$ or $\sigma \geq 22.112$ under model M1 and for $\sigma = 2.569$ or $\sigma \geq 20.549$ under model M2.
On the basis of this assessment, the $\text{MSPEB}_{\text{P}}$ criterion outperforms the $\text{MSPEB}_{\text{D}}$ and $\text{MSPEB}_{\text{Y}}$ criteria for these values of $\sigma$ under models M1 and M2 and the $\text{MSPE}_{\text{Y}}$ criterion outperforms the $\text{MSPE}_{\text{D}}$ and $\text{MSPE}_{\text{P}}$ criteria for the other values of $\sigma$.
The MINR and RC of the true model and a model selected by any one of the MSPE criteria in the TCL method are generally comparable to or even higher than those of the true model and a model selected by the same criterion in the LCT method.
Together with the information of estimated numbers of clusters in Table \ref{tab:simu_c}, the MSPEB criteria outperform the corresponding MSPE criteria in terms of the MINR in the TCL method for $\sigma \leq 5.528$ or $\sigma \geq 22.112$ under model M1 and for $\sigma \geq 20.549$ under model M2.
Especially, the $\text{MSPEB}_{\text{P}}$ criterion outperforms the $\text{MSPEB}_{\text{D}}$ and $\text{MSPEB}_{\text{Y}}$ criteria for these values of $\sigma$ under models M1 and M2 and the $\text{MSPE}_{\text{Y}}$ criterion outperforms the $\text{MSPE}_{\text{D}}$ and $\text{MSPE}_{\text{P}}$ criteria for the other values of $\sigma$.
Compared to any one of the MSPE criteria in the TCL method, the corresponding MSPEB criterion has a higher RC for $\sigma \leq 11.056$ under model M1 and for $\sigma \leq 5.137$ under model M2.
Except for $\sigma = 11.056$ under model M1, the $\text{MSPEB}_{\text{P}}$ criterion outperforms the $\text{MSPEB}_{\text{D}}$ and $\text{MSPEB}_{\text{Y}}$ criteria in terms of the RC.
Moreover, the $\text{MSPE}_{\text{Y}}$ criterion produces the highest RC for the other values of $\sigma$.
\begin{table}[htbp]
  \centering
  \caption{The mean squared errors of estimated robot strengths from the WMPR model and selected optimal models in both the LCT and TCL methods under models M1 and M2.}
  \resizebox{\columnwidth}{!}{%
    \begin{tabular}{crrrrrrrrrrrr}
    \toprule
    Model & \multicolumn{1}{c}{$\sigma$} & \multicolumn{1}{c}{WMPR}  & \multicolumn{3}{c}{LCT method} & & \multicolumn{6}{c}{TCL method}\\
    \cline{4-6} \cline{8-13}
         &   &   & \multicolumn{1}{c}{$\textsc{mspe}_{\textsc{d}}$}  & \multicolumn{1}{c}{$\textsc{mspe}_{\textsc{p}}$} &  \multicolumn{1}{c}{$\textsc{mspe}_{\textsc{y}}$} &       & \multicolumn{1}{c}{$\textsc{mspe}_{\textsc{d}}$} & \multicolumn{1}{c}{$\textsc{mspeb}_{\textsc{d}}$} & \multicolumn{1}{c}{$\textsc{mspe}_{\textsc{p}}$} & \multicolumn{1}{c}{$\textsc{mspeb}_{\textsc{p}}$} & \multicolumn{1}{c}{$\textsc{mspe}_{\textsc{y}}$} & \multicolumn{1}{c}{$\textsc{mspeb}_{\textsc{y}}$} \\
         \midrule
         M1 & 2.764   &96.690     & 65.529  & 73.692 & 82.870  &       & 60.595 & 47.971 & 68.194 & 39.004 & 79.212 & 52.500 \\
          & 5.528   &386.758     & 344.336  & 353.795  & 368.416 &       & 303.266 & 257.720 & 321.996 & 229.984 & 354.855 & 291.340 \\
     & 11.056  &1547.033      & 1608.947  & 1591.066  & 1577.577  &       & 1588.265 & 1647.903 & 1576.648 & 1657.808 & 1573.168 & 1599.839 \\
          & 22.112  &6188.132      & 5983.565 & 6037.026 & 6072.589  &       & 5677.451 & 5195.265 & 5817.771 & 5145.647 & 5964.496 & 5562.063 \\
          & 44.223   &24752.527     & 22723.638 & 23151.075  & 23501.749  &       & 21053.862 & 17324.722 & 21898.330 & 16660.093 & 22884.581 & 19397.953 \\
          \midrule \midrule
       M2   & 2.569  &  85.063    & 72.497 & 75.350 & 79.880 &       & 63.613 & 61.119 & 67.955 & 58.758 & 76.556 & 59.334 \\
          & 5.137  &340.250      & 368.003  & 358.792  & 352.319  &       & 365.250 & 393.390 & 358.392 & 396.763 & 351.848 & 364.888 \\
     & 10.275 &  1360.710     & 1408.038  & 1397.063  & 1387.490 &       & 1391.700 & 1446.551 & 1383.382 & 1459.470 & 1379.467 & 1400.618 \\
          & 20.549   &5442.841     & 5275.136   & 5319.028  & 5352.931  &       & 5022.987 & 4673.683 & 5137.169 & 4619.494 & 5246.074 & 4926.585 \\
          & 41.098    &21776.030    & 20025.404  & 20417.138 & 20728.255  &       & 18601.295 & 15644.076 & 19293.720 & 14995.979 & 20118.878 & 17202.609 \\
               \bottomrule
    \end{tabular}%
    }
  \label{tab:simu_mse}%
\end{table}%

\begin{table}[htbp]
\tiny
  \centering
  \caption{The averages of 5000 MINR's of the true model and the WMPR model and those of the true model and selected optimal models in both the LCT and TCL methods under models M1 and M2.}
  \resizebox{\columnwidth}{!}{%
    \begin{tabular}{crrrrrrrrrrrr}
    \toprule
    Model & \multicolumn{1}{c}{$\sigma$} & \multicolumn{1}{c}{WMPR}  & \multicolumn{3}{c}{LCT method} & & \multicolumn{6}{c}{TCL method}\\
    \cline{4-6} \cline{8-13}
         &   &   & \multicolumn{1}{c}{$\textsc{mspe}_{\textsc{d}}$}  & \multicolumn{1}{c}{$\textsc{mspe}_{\textsc{p}}$} &  \multicolumn{1}{c}{$\textsc{mspe}_{\textsc{y}}$} &       & \multicolumn{1}{c}{$\textsc{mspe}_{\textsc{d}}$} & \multicolumn{1}{c}{$\textsc{mspeb}_{\textsc{d}}$} & \multicolumn{1}{c}{$\textsc{mspe}_{\textsc{p}}$} & \multicolumn{1}{c}{$\textsc{mspeb}_{\textsc{p}}$} & \multicolumn{1}{c}{$\textsc{mspe}_{\textsc{y}}$} & \multicolumn{1}{c}{$\textsc{mspeb}_{\textsc{y}}$} \\
         \midrule
         M1 & 2.764 & 100.0\% &    100.0\% & 100.0\% & 100.0\% &       & 100.0\% & 100.0\% & 100.0\% & 100.0\% & 100.0\% & 100.0\% \\
          & 5.528 &   96.4\%&     96.1\%  & 96.2\%  & 96.3\% &       & 96.8\% & 97.0\% & 96.7\% & 97.2\% & 96.5\% & 96.8\% \\
     & 11.056 &   73.5\%&   72.2\% & 72.5\% & 72.9\% &       & 72.3\% & 70.7\% & 72.6\% & 70.5\% & 72.9\% & 72.1\% \\
          & 22.112 &    47.3\%&    47.1\%  & 47.2\%  & 47.2\%  &       & 47.7\% & 48.0\% & 47.6\% & 47.8\% & 47.4\% & 47.7\% \\
          & 44.223 &   30.9\%&     31.1\%  & 31.0\%  & 31.0\% &       & 31.8\% & 32.6\% & 31.5\% & 32.7\% & 31.2\% & 32.0\% \\
\midrule \midrule
       M2   & 2.569 & 94.4\%&  93.5\% & 94.0\%  & 94.2\% &       & 94.4\% & 93.8\% & 94.6\% & 93.6\% & 94.5\% & 94.9\% \\
          & 5.137 &   75.1\%&     73.0\% & 73.7\%  & 74.1\%  &       & 73.0\% & 71.2\% & 73.5\% & 71.0\% & 74.0\% & 72.8\% \\
     & 10.275 &     48.5\%&   47.3\%  & 47.6\% & 47.9\%  &       & 47.4\% & 46.1\% & 47.7\% & 45.9\% & 47.9\% & 47.2\% \\
          & 20.549 &     28.3\%&   28.4\%  & 28.4\% & 28.4\%  &       & 28.8\% & 29.0\% & 28.8\% & 29.0\% & 28.6\% & 28.8\% \\
          & 41.098 &     16.9\%&   17.2\% & 17.1\%  & 17.0\%  &       & 17.6\% & 18.3\% & 17.5\% & 18.4\% & 17.3\% & 17.8\% \\
       \bottomrule
    \end{tabular}%
    }
  \label{tab:simu_minr}%
\end{table}%

\begin{table}[htbp]
\tiny
  \centering
  \caption{The averages of 5000 RC's of the true model and the WMPR model and those of the true model and selected optimal models in both the LCT and TCL methods under models M1 and M2.}
  \resizebox{\columnwidth}{!}{%
    \begin{tabular}{crrrrrrrrrrrr}
    \toprule
    Model & \multicolumn{1}{c}{$\sigma$} & \multicolumn{1}{c}{WMPR}  & \multicolumn{3}{c}{LCT method} & & \multicolumn{6}{c}{TCL method}\\
    \cline{4-6} \cline{8-13}
         &   &   & \multicolumn{1}{c}{$\textsc{mspe}_{\textsc{d}}$}  & \multicolumn{1}{c}{$\textsc{mspe}_{\textsc{p}}$} &  \multicolumn{1}{c}{$\textsc{mspe}_{\textsc{y}}$} &       & \multicolumn{1}{c}{$\textsc{mspe}_{\textsc{d}}$} & \multicolumn{1}{c}{$\textsc{mspeb}_{\textsc{d}}$} & \multicolumn{1}{c}{$\textsc{mspe}_{\textsc{p}}$} & \multicolumn{1}{c}{$\textsc{mspeb}_{\textsc{p}}$} & \multicolumn{1}{c}{$\textsc{mspe}_{\textsc{y}}$} & \multicolumn{1}{c}{$\textsc{mspeb}_{\textsc{y}}$} \\
         \midrule
     M1    & 2.764 & 70.7\%&    83.9\%  & 81.9\%  & 79.6\%  &       & 84.6\% & 88.1\% & 82.6\% & 90.5\% & 80.2\% & 87.0\% \\
    & 5.528 & 70.6\%&  82.2\% &  80.6\%  & 78.8\%  &       & 83.4\% & 87.3\% & 81.7\% & 89.5\% & 79.4\% & 85.2\% \\
     & 11.056 & 67.0\%&   71.7\%  & 71.5\%  & 71.3\%  &       & 71.5\% & 71.8\% & 71.4\% & 72.0\% & 71.3\% & 72.3\% \\
   & 22.112 & 58.0\%& 58.6\%  & 58.7\% & 58.9\%  &    &    58.6\% & 57.8\% & 58.8\% & 57.6\% & 58.8\% & 58.6\% \\
    & 44.223 & 48.5\%& 47.5\% & 47.7\% & 47.9\%  &       & 47.6\% & 46.5\% & 47.8\% & 46.1\% & 48.0\% & 47.1\% \\
\midrule \midrule
M2 & 2.569 & 80.8\%  & 90.7\%  & 88.9\%  & 87.1\%  &       & 91.6\% & 93.3\% & 89.9\% & 94.1\% & 87.6\% & 92.4\% \\
    & 5.137 & 79.0\%&  83.4\%  & 83.0\%  & 82.7\%  &       & 83.4\% & 83.7\% & 83.2\% & 84.0\% & 82.9\% & 84.1\% \\
     & 10.275 & 73.2\%&  73.2\%  & 73.4\% & 73.7\%  &       & 73.0\% & 71.6\% & 73.4\% & 71.4\% & 73.6\% & 73.0\% \\
    & 20.549 & 63.9\%& 60.8\%  & 61.3\% & 61.8\%  &       & 60.9\% & 58.4\% & 61.3\% & 57.6\% & 61.8\% & 59.9\% \\
    & 41.098&  54.3\%& 49.9\% & 50.4\% &  50.8\%  &       & 49.9\% & 47.0\% & 50.3\% & 46.0\% & 50.8\% & 48.2\% \\
       \bottomrule
    \end{tabular}%
    }
  \label{tab:simu_rc}%
\end{table}%

In Tables \ref{tab:simu_msped} -- \ref{tab:simu_mspey}, the MSPE measures of selected optimal models are underestimated in both the LCT and TCL methods.
This underestimation becomes severe as $\sigma$ increases.
For selected optimal models with the minimal MSPE measures, the closest estimates are further found in the TCL method.
Moreover, the $\text{MSPE}_{\text{D}}$ of a model selected by any one of the MSPEB criteria is smaller than that selected by the corresponding MSPE criterion for $\sigma \leq 5.528$ under model M1 and for $\sigma = 2.569$ under model M2.
Compared to a model selected by any one of the MSPE criteria, a model selected by the corresponding MSPEB criteria has smaller $\text{MSPE}_{\text{P}}$ and $\text{MSPE}_{\text{Y}}$ for $\sigma \leq 5.528$ or $\sigma \geq 22.112$ under model M1 and for $\sigma = 2.569$ or $\sigma \geq 20.549$ under model M2.
Especially, a model selected by the $\text{MSPEB}_{\text{P}}$ criterion has the minimal $\text{MSPE}_{\text{Y}}$ and the closest $\text{MSPE}_{\text{Y}}$ estimate.
A model selected by the $\text{MSPE}_{\text{Y}}$ criterion also has this advantage for the other values of $\sigma$.

\begin{table}[htb]
\tiny
  \centering
  \caption{The $\text{MSPE}_{\text{D}}$'s and the averages of their 5000 estimates (within parenthesis) of the WMPR model and selected optimal models in both the LCT and TCL methods under models M1 and M2.}
  \resizebox{\columnwidth}{!}{%
    \begin{tabular}{crrrrrrrrrrrr}
    \toprule
    Model & \multicolumn{1}{c}{$\sigma$} & \multicolumn{1}{c}{WMPR}  & \multicolumn{3}{c}{LCT method} & & \multicolumn{6}{c}{TCL method}\\
    \cline{4-6} \cline{8-13}
         &   &   & \multicolumn{1}{c}{$\textsc{mspe}_{\textsc{d}}$}  & \multicolumn{1}{c}{$\textsc{mspe}_{\textsc{p}}$} &  \multicolumn{1}{c}{$\textsc{mspe}_{\textsc{y}}$} &       & \multicolumn{1}{c}{$\textsc{mspe}_{\textsc{d}}$} & \multicolumn{1}{c}{$\textsc{mspeb}_{\textsc{d}}$} & \multicolumn{1}{c}{$\textsc{mspe}_{\textsc{p}}$} & \multicolumn{1}{c}{$\textsc{mspeb}_{\textsc{p}}$} & \multicolumn{1}{c}{$\textsc{mspe}_{\textsc{y}}$} & \multicolumn{1}{c}{$\textsc{mspeb}_{\textsc{y}}$} \\
         \midrule
       M1   & 2.764 & 0.101&  0.100 & 0.100  & 0.100  &       & 0.100 & 0.099 & 0.100 & 0.099 & 0.101 & 0.100 \\
          &       &   (0.102)&     (0.041)  & (0.048)  & (0.055)  &       & (0.039) & (0.043) & (0.046) & (0.056) & (0.055) & (0.061) \\
          & 5.528 &   0.121&     0.121 & 0.120 & 0.120 &       & 0.121 & 0.120 & 0.121 & 0.120 & 0.120 & 0.120 \\
          &       &    (0.142)&    (0.055)  & (0.063)  & (0.070) &       & (0.054) & (0.058) & (0.061) & (0.072) & (0.070) & (0.075) \\
     & 11.056 &   0.213&    0.217  & 0.216  & 0.215  &       & 0.216 & 0.221 & 0.216 & 0.223 & 0.215 & 0.218 \\
          &       &    (0.252)&    (0.104)  & (0.114)  & (0.120)  &       & (0.103) & (0.111) & (0.113) & (0.124) & (0.121) & (0.126) \\
          & 22.112 &   0.340&     0.343  & 0.342 & 0.341  &       & 0.342 & 0.347 & 0.341 & 0.349 & 0.341 & 0.344 \\
          &  & (0.376)& (0.162)  & (0.175)  & (0.181)  & & (0.161) & (0.172) & (0.174) & (0.191) & (0.182) & (0.190) \\
          & 44.223 &    0.439  &  0.440  & 0.440  & 0.439 &       & 0.440 & 0.442 & 0.439 & 0.443 & 0.439 & 0.441 \\
         & &  (0.459)& (0.203) & (0.217)  & (0.223)  &  & (0.202) & (0.215) & (0.216) & (0.238) & (0.224) & (0.235) \\
\midrule \midrule
M2 & 2.569 &  0.062&   0.060 & 0.060 & 0.060 &       & 0.060 & 0.060 & 0.060 & 0.060 & 0.060 & 0.059 \\
          &&(0.072)&(0.027)&(0.033)&(0.039)&&(0.026)&(0.029)&(0.031)&(0.037)&(0.039)&(0.042) \\
          & 5.137 &  0.114&      0.116 & 0.115 & 0.114 &       & 0.116 & 0.119 & 0.115 & 0.120 & 0.114 & 0.116 \\
         &&(0.138)&(0.055)&(0.063)&(0.069)&&(0.054)&(0.059)&(0.061)&(0.069)&(0.069)&(0.073) \\
     & 10.275 &    0.214&    0.217 & 0.216  & 0.215  &       & 0.217 & 0.222 & 0.216 & 0.224 & 0.215 & 0.218 \\
         &&(0.250)&(0.105)&(0.115)&(0.121)&&(0.103)&(0.111)&(0.113)&(0.126)&(0.122)&(0.128) \\
          & 20.549 &  0.338&      0.340  & 0.339  & 0.339  &       & 0.340 & 0.345 & 0.340 & 0.346 & 0.338 & 0.342 \\
         &  &(0.373)   &(0.163)&(0.176)&(0.181)&&(0.161)&(0.171)&(0.174)&(0.190)&(0.182)&(0.189) \\
          & 41.098 &   0.436&    0.437  & 0.437 & 0.436  &       & 0.437 & 0.439 & 0.436 & 0.440 & 0.436 & 0.438 \\
&&(0.455)&(0.202)&(0.217)&(0.222)&&(0.201)&(0.213)&(0.215)&(0.236)&(0.224)&(0.234) \\
       \bottomrule
    \end{tabular}%
    }
  \label{tab:simu_msped}%
\end{table}%

\begin{table}[htbp]
\tiny
  \centering
  \caption{The $\text{MSPE}_{\text{P}}$'s and the averages of their 5000 estimates (within parenthesis) of the WMPR model and selected optimal models in both the LCT and TCL methods under models M1 and M2.}
    \resizebox{\columnwidth}{!}{%
    \begin{tabular}{crrrrrrrrrrrr}
    \toprule
    Model & \multicolumn{1}{c}{$\sigma$} & \multicolumn{1}{c}{WMPR}  & \multicolumn{3}{c}{LCT method} & & \multicolumn{6}{c}{TCL method}\\
    \cline{4-6} \cline{8-13}
         &   &   & \multicolumn{1}{c}{$\textsc{mspe}_{\textsc{d}}$}  & \multicolumn{1}{c}{$\textsc{mspe}_{\textsc{p}}$} &  \multicolumn{1}{c}{$\textsc{mspe}_{\textsc{y}}$} &       & \multicolumn{1}{c}{$\textsc{mspe}_{\textsc{d}}$} & \multicolumn{1}{c}{$\textsc{mspeb}_{\textsc{d}}$} & \multicolumn{1}{c}{$\textsc{mspe}_{\textsc{p}}$} & \multicolumn{1}{c}{$\textsc{mspeb}_{\textsc{p}}$} & \multicolumn{1}{c}{$\textsc{mspe}_{\textsc{y}}$} & \multicolumn{1}{c}{$\textsc{mspeb}_{\textsc{y}}$} \\
         \midrule
       M1     & 2.764 & 0.071&  0.066 & 0.068 & 0.069 &       & 0.066 & 0.064 & 0.068 & 0.062 & 0.069 & 0.065 \\
        &  & (0.081)& (0.036)  & (0.033) & (0.037)  & & (0.035) & (0.036) & (0.032) & (0.036) & (0.037) & (0.039) \\
    & 5.528 &  0.089& 0.087  & 0.088  & 0.088  &       & 0.086 & 0.083 & 0.087 & 0.082 & 0.088 & 0.085 \\
            & &  (0.112)& (0.048)  & (0.045)  & (0.049)  &  & (0.047) & (0.049) & (0.044) & (0.049) & (0.049) & (0.051) \\
     & 11.056 & 0.160&  0.162 & 0.161 & 0.161 &       & 0.162 & 0.163 & 0.161 & 0.163 & 0.161 & 0.161 \\
          & & (0.200)&  (0.084)  & (0.081)  & (0.084)  & & (0.084) & (0.088) & (0.080) & (0.086) & (0.085) & (0.087) \\
   & 22.112 & 0.250& 0.249  & 0.249 & 0.249 &       & 0.248 & 0.246 & 0.249 & 0.246 & 0.249 & 0.247 \\
           & &  (0.298)& (0.125)  & (0.122)  & (0.125)  & & (0.125) & (0.131) & (0.121) & (0.130) & (0.125) & (0.129) \\
    & 44.223 & 0.315& 0.310  & 0.311  & 0.312 &       & 0.310 & 0.302 & 0.311 & 0.301 & 0.312 & 0.306 \\
       & &  (0.362)& (0.152)  & (0.149)  & (0.151)  & & (0.153) & (0.158) & (0.149) & (0.158) & (0.152) & (0.157) \\
\midrule \midrule
M2 & 2.569 & 0.046& 0.044  & 0.044 & 0.045  &       & 0.043 & 0.043 & 0.044 & 0.043 & 0.045 & 0.043 \\
          &   & (0.057)& (0.026)  & (0.024)  & (0.027) & & (0.025) & (0.027) & (0.023) & (0.025) & (0.027) & (0.029) \\
    & 5.137 & 0.086&  0.087  & 0.086 & 0.086  &       & 0.087 & 0.088 & 0.086 & 0.088 & 0.086 & 0.086 \\
            &  & (0.109)&  (0.048)  & (0.045)  & (0.049)  &  & (0.047) & (0.050) & (0.044) & (0.049) & (0.049) & (0.051) \\
     & 10.275 & 0.161& 0.161& 0.161 & 0.161 &       & 0.161 & 0.163 & 0.161 & 0.163 & 0.161 & 0.161 \\
           &  &  (0.198)& (0.085) & (0.082)  & (0.085) &  & (0.085) & (0.088) & (0.081) & (0.087) & (0.085) & (0.088) \\
    & 20.549 & 0.248& 0.247  & 0.247  & 0.247 &       & 0.246 & 0.244 & 0.247 & 0.244 & 0.247 & 0.245 \\
            &   &(0.295)& (0.125)  & (0.122)& (0.125)  & & (0.125) & (0.130) & (0.121) & (0.129) & (0.125) & (0.129) \\
    & 41.098&  0.313& 0.308  & 0.309  & 0.310  &       & 0.307 & 0.301 & 0.309 & 0.299 & 0.310 & 0.304 \\
         &   & (0.358)&(0.151) & (0.148)  & (0.151)  && (0.152) & (0.157) & (0.148) & (0.157) & (0.152) & (0.156) \\
       \bottomrule
    \end{tabular}%
    }
  \label{tab:simu_mspep}%
\end{table}%

\begin{table}[htbp]
  \centering
  \caption{The $\text{MSPE}_{\text{Y}}$'s and the averages of their 5000 estimates (within parenthesis) of the WMPR model and selected optimal models in both the LCT and TCL methods under models M1 and M2.}
    \resizebox{\columnwidth}{!}{%
   \begin{tabular}{crrrrrrrrrrrr}
    \toprule
    Model & \multicolumn{1}{c}{$\sigma$} & \multicolumn{1}{c}{WMPR}  & \multicolumn{3}{c}{LCT method} & & \multicolumn{6}{c}{TCL method}\\
    \cline{4-6} \cline{8-13}
         &   &   & \multicolumn{1}{c}{$\textsc{mspe}_{\textsc{d}}$}  & \multicolumn{1}{c}{$\textsc{mspe}_{\textsc{p}}$} &  \multicolumn{1}{c}{$\textsc{mspe}_{\textsc{y}}$} &       & \multicolumn{1}{c}{$\textsc{mspe}_{\textsc{d}}$} & \multicolumn{1}{c}{$\textsc{mspeb}_{\textsc{d}}$} & \multicolumn{1}{c}{$\textsc{mspe}_{\textsc{p}}$} & \multicolumn{1}{c}{$\textsc{mspeb}_{\textsc{p}}$} & \multicolumn{1}{c}{$\textsc{mspe}_{\textsc{y}}$} & \multicolumn{1}{c}{$\textsc{mspeb}_{\textsc{y}}$} \\
         \midrule
       M1  & 2.764 & 12.152&   11.091  & 11.410  & 11.782 &       & 11.112 & 10.606 & 11.394 & 10.143 & 11.781 & 10.925 \\
& &(18.897)& (5.261)&(5.069)&(4.644)& &(5.257)&(5.608)&(5.084)&(5.888)&(4.645)&(5.133) \\
          & 5.528 &   48.608&     48.165  & 48.138 & 48.432  &       & 46.769 & 45.346 & 47.294 & 44.115 & 48.151 & 46.502 \\
& & (75.589)&(20.667)&(19.683)&(18.166)& &(20.626)&(22.484)&(19.845)&(23.677)&(18.181)&(19.871) \\
     & 11.056 &  194.433&      201.063  & 198.969 & 197.359  &       & 200.611 & 208.177 & 199.000 & 208.606 & 197.107 & 200.944 \\
 & &(302.355)& (72.395)&(69.716)&(66.651)& &(73.846)&(83.668)&(71.008)&(83.214)&(67.053)&(70.688) \\
          & 22.112 &    777.732&    780.588  & 779.978 & 779.469 &       & 777.174 & 776.823 & 777.966 & 777.023 & 778.565 & 777.880 \\
 & & (1209.420)&(266.464)&(258.957)&(251.761)& &(274.191)&(300.087)&(264.440)&(303.808)&(254.748)&(268.698) \\
          & 44.223 &    3110.927&    3063.919 & 3075.108 & 3084.111  &       & 3051.233 & 2971.004 & 3066.674 & 2951.641 & 3083.676 & 3018.019 \\
& & (4837.680)& (1027.002)&(1004.256)&(981.964)& &(1056.786)&(1136.749)&(1025.637)&(1150.639)&(995.138)&(1052.745) \\
\midrule \midrule
M2  & 2.569 & 10.506&   10.545  & 10.412  & 10.411  &       & 10.176 & 10.333 & 10.192 & 10.307 & 10.349 & 9.896 \\
 & &(16.182)& (5.692)&(5.114)&(4.443)& &(5.563)&(6.295)&(5.115)&(6.535)&(4.442)&(5.018) \\
          & 5.137 &   42.025&     44.312 & 43.461  & 42.740  &       & 44.243 & 46.575 & 43.539 & 46.768 & 42.637 & 43.920 \\
 &  &(64.728)&(18.436)&(17.434)&(16.131)& &(18.765)&(21.493)&(17.847)&(21.712)&(16.175)&(17.513) \\
          & 10.275 &    168.102&    172.790  & 171.344  & 170.052  &       & 172.653 & 178.761 & 171.302 & 179.692 & 169.591 & 173.000 \\
& & (258.911)&(63.034)&(60.892)&(58.129)& &(64.607)&(72.424)&(62.056)&(73.029)&(58.446)&(61.989) \\
          & 20.549 &    672.407&    673.753 & 672.921 & 671.913  &       & 670.860 & 670.643 & 670.683 & 670.228 & 670.457 & 670.691 \\
& &(1035.643)& (231.241)&(224.506)&(218.217)& &(236.383)&(257.375)&(228.701)&(261.420)&(220.202)&(232.390) \\
          & 41.098 &   2689.630&     2641.107  & 2650.044 & 2657.535 &       & 2630.466 & 2569.382 & 2643.275 & 2554.320 & 2657.237 & 2606.379 \\
 & & (4142.571)&(888.842)&(868.555)&(848.412)& &(908.573)&(969.931)&(884.434)&(984.930)&(857.026)&(904.398) \\
       \bottomrule
    \end{tabular}%
    }
  \label{tab:simu_mspey}%
\end{table}%
\end{subsection}
\end{section}

\begin{section}{Concluding Remarks and Future Challenges} \label{sec:concluding}
Compared to existing methods, a more effective method is developed to identify clusters of robot strengths in an application to qualification stage data from the 2018 and 2019 FRC Houston and Detroit championships.
In our method, a classification system is first proposed to determine WMPRC candidate models.
The designed non-hierarchical classification is found to be particularly useful for examining possible misclassification of robots in each hierarchical classification step.
The MSPE and MSPEB criteria are further established to select optimal models from these candidate models. 
By coherently assembling the two parts, a systematic procedure is, thus, presented to estimate the number of clusters, clusters of robots and robot strengths.
In addition to these achievements, the MINR and modified RC are provided, respectively, to measure the nested relation between clusters from any two models and monotonic association between robot strengths from any two models.

With the same MSPE criterion for model selection, the conducted simulation shows that our classification system can reduce the MSE of estimated robot strengths in the LCT method.
The MINR and RC in the TCL method are further found to be comparable to or even higher than those in the LCT method.
For selected optimal models with the minimal MSPE measures, which are underestimated in both the methods, the TCL method tends to produce the closest MSPE estimates.
Especially, models selected by the MSPEB criteria are shown to have better performance than those selected by the corresponding MSPE criteria when the proportion of sufficiently separated robot strengths is higher or lower than some threshold value.
Moreover, the MSPE measures and their estimates of a selected optimal model are larger and smaller than the corresponding MSPE measures and estimates of the true WMPRC model.
This underestimation becomes more severe as the standard deviation of the error increases.
The established MSPE or MSPEB criteria are still useful for selecting optimal models, albeit MSPE estimates are inappropriate to serve as assessments in the proposed classification system.
It is notable that a pairwise fusion method, which is introduced in Remark \ref{rem:fusion}, is superior to the developed method for strongly separated clusters of robot strengths.
However, our method has been demonstrated to perform better on data similar to those of the qualification stage of the FRC championships and is recommended for adoption by FRC teams to evaluate potential alliance partners in the playoff stage.

Regarding the model structure, the WMPRC model describes the linear dependence of the difference in scores on robot strengths.
Compared to most of the pairwise differences between different robot strengths, the standard deviation of the error is considerably large in qualification stage data from the FRC championships.
Much of the variation may be attributed to the current scoring system in the FRC and/or the degree of model misspecification.
A possible way to reduce the variation of the scoring system is to increase the number of matches in the qualification stage for model training.
Although robot strengths in the WMPRC model can be generalized to random robot strengths, robot-specific-effects are negligible  on the variation of the error in our data analysis. 
The results are not shown due to space limitations.
For the model misspecification problem, another avenue is to borrow the neural network or deep learning framework \cite[cf.][]{rosenblatt1961principles, L2015} for the relation between the difference in scores and robot strengths with a monotonicity constraint on the link function.
Our method should be applicable for parameter estimation in this model approximation.
We note that the variation in scores may also be caused by some factors such as robot characteristics, penalty scores and environmental factors.
A more thorough study warrants future research.
\end{section}

\beginsupplement

\begin{supplement}
\textbf{Supplementary data and codes.} The data and R codes used in this article are available at \url{https://github.com/d09948011/Identifying-clusters}.
\end{supplement}

\begin{supplement} 
\textbf{Data analysis and simulation results.} The supplementary tables referenced in Remark \ref{rem:fusion}.
~
\begin{table}[htbp]
  \tiny
  \centering
  \caption{The numbers of clusters in models estimated by the criterion in (\ref{cpfusion}) as well as BIC, the $\text{PCP}$, $\text{MSPE}_{\text{P}}$ and $\text{MSPE}_{\text{Y}}$ estimates of these estimated models and the numbers of correct predictions of semifinalists, finalists and champions in the playoffs for the 2018 and 2019 FRC championships.}
        \resizebox{!}{3.73cm}{%
    \begin{tabular}{lllrrrrrrrr}
    \toprule    
     Season & Tournament & Division & $\widehat{c}$ & \multicolumn{1}{c}{$\widehat{\textsc{pcp}}$} & \multicolumn{1}{c}{$\widehat{\textsc{mspe}}_{\textsc{p}}$} & \multicolumn{1}{c}{$\widehat{\textsc{mspe}}_{\textsc{y}}$} & \multicolumn{1}{c}{Semifinalists}  &\multicolumn{1}{c}{Finalists} & \multicolumn{1}{c}{Champions}  \\
     \midrule
    2018  & Houston & Carver & 9     & 82.5\% & 0.126 & 8896.8 & 3     & 0     & 1 \\
          &       & Galileo & 10    & 86.0\% & 0.115 & 9180.4 & 3     & 2     & 0 \\
          &       & Hopper & 8     & 84.2\% & 0.114 & 9326.7 & 3     & 2     & 0 \\
          &       & Newton & 7     & 89.7\% & 0.083 & 7148.6 & 3     & 2     & 0 \\
          &       & Roebling & 16    & 86.2\% & 0.118 & 7901.9 & 2     & 1     & 0 \\
          &       & Turing & 8     & 89.2\% & 0.083 & 9046.2 & 3     & 1     & 0 \\
                             \cline{2-10}
          & Detroit & Archimedes & 17    & 78.1\% & 0.139 & 8574.4 & 3     & 0     & 1 \\
          &       & Carson & 10    & 76.8\% & 0.139 & 9831.4 & 4     & 1     & 1 \\
          &       & Curie & 6     & 89.3\% & 0.094 & 6853.5 & 2     & 1     & 1 \\
          &       & Daly  & 10    & 89.0\% & 0.082 & 8195.1 & 3     & 2     & 0 \\
          &       & Darwin & 8     & 83.5\% & 0.128 & 9815.1 & 1     & 2     & 0 \\
          &       & Tesla & 11    & 86.6\% & 0.116 & 8557.8 & 2     & 1     & 1 \\
             \midrule
    Average &       &       & 10    & 85.1\% & 0.111 & 8610.7 & 2.7   & 1.3   & 0.4 \\
        \midrule
        \midrule
    2019  & Houston & Carver & 9     & 81.6\% & 0.128 & 186.4 & 4     & 1     & 1 \\
          &       & Galileo & 10    & 89.7\% & 0.095 & 124.5 & 2     & 2     & 0 \\
          &       & Hopper & 11    & 83.3\% & 0.101 & 112.2 & 3     & 1     & 1 \\
          &       & Newton & 12    & 82.1\% & 0.121 & 173.7 & 3     & 1     & 1 \\
          &       & Roebling & 9     & 77.2\% & 0.145 & 164.4 & 2     & 1     & 0 \\
          &       & Turing & 7     & 82.7\% & 0.115 & 153.3 & 1     & 1     & 1 \\
                                       \cline{2-10}
          & Detroit & Archimedes & 7     & 78.9\% & 0.150 & 227.3 & 2     & 1     & 1 \\
          &       & Carson & 13    & 80.7\% & 0.122 & 111.2 & 1     & 1     & 0 \\
          &       & Curie & 7     & 82.5\% & 0.118 & 175.3 & 4     & 1     & 1 \\
          &       & Daly  & 5     & 74.6\% & 0.152 & 189.9 & 2     & 1     & 0 \\
          &       & Darwin & 5     & 75.0\% & 0.160 & 148.1 & 3     & 1     & 0 \\
          &       & Tesla & 7     & 81.6\% & 0.135 & 178.4 & 2     & 1     & 1 \\
    \midrule
    Average &       &       & 8.5   & 80.8\% & 0.128 & 162.1 & 2.4   & 1.1   & 0.6 \\
     \bottomrule
    \end{tabular}%
    }
  \label{tab:fusion}%
\end{table}%
~\\
\begin{table}[htbp]
\tiny
  \centering
  \caption{The means of 5000 estimated numbers of clusters, mean squared errors of estimated robot strengths, averages of 5000 MINR's between the true and estimated clusters, averages of 5000 RC's between the true and estimated robot strengths and $\text{MSPE}_{\text{D}}$, $\text{MSPE}_{\text{P}}$ and $\text{MSPE}_{\text{Y}}$ of models estimated by the criterion in (\ref{cpfusion}).}
    \begin{tabular}{ccrrrrrrr}
    \toprule
     \multicolumn{1}{c}{Model} & \multicolumn{1}{c}{$\sigma$} & \multicolumn{1}{c}{$\widehat{c}$} & \multicolumn{1}{c}{MSE} & \multicolumn{1}{c}{$\textsc{minr}$} & \multicolumn{1}{c}{$\textsc{rc}$} & \multicolumn{1}{c}{$\textsc{mspe}_{\textsc{d}}$} & \multicolumn{1}{c}{$\textsc{mspe}_{\textsc{p}}$} & \multicolumn{1}{c}{$\textsc{mspe}_{\textsc{y}}$}\\
     \midrule
    \multicolumn{1}{c}{M1} & 2.764 & 4.3   & 3.195 & 100.0\% & 99.9\% & 0.093 & 0.051 & 7.899 \\
          & 5.528 & 9.0   & 120.898 & 98.3\% & 96.1\% & 0.119 & 0.073 & 38.235 \\
          & 11.056 & 7.9   & 2205.229 & 61.9\% & 66.9\% & 0.255 & 0.173 & 242.470 \\
          & 22.112 & 7.1   & 5439.416 & 46.0\% & 53.8\% & 0.363 & 0.245 & 788.623 \\
          & 44.223 & 6.1   & 15676.612 & 34.1\% & 42.5\% & 0.452 & 0.292 & 2855.011 \\
\midrule
        \midrule
    \multicolumn{1}{c}{M2} & 2.569 & 18.0  & 47.067 & 94.1\% & 95.8\% & 0.062 & 0.042 & 9.632 \\
          & 5.137 & 17.0  & 450.986 & 67.2\% & 82.5\% & 0.129 & 0.092 & 49.023 \\
          & 10.275 & 8.3   & 1789.001 & 38.7\% & 60.1\% & 0.262 & 0.171 & 208.331 \\
          & 20.549 & 7.1   & 4747.166 & 30.0\% & 49.3\% & 0.368 & 0.244 & 685.456 \\
          & 41.098 & 5.9   & 13005.827 & 23.6\% & 37.2\% & 0.453 & 0.286 & 2443.311 \\
           \bottomrule
    \end{tabular}%
  \label{tab:fusion_simu}%
\end{table}%
\end{supplement} 

\bibliographystyle{imsart-nameyear} 
\bibliography{reference}       

\end{document}